\documentclass[11pt,openright,twoside]{report}

\usepackage{geometry}
\usepackage{setspace} %to set 1.5 spacing in text
\usepackage{graphicx}
\usepackage[font=footnotesize,labelfont=bf]{caption}%to costumize captions
\usepackage{subfig} %put this below caption package to avoid redefinition problems
\usepackage{color} 
\usepackage{amsmath}
\usepackage{amssymb}
\usepackage{verbatim}

\begin{document}

\title{\textbf{Kinetic Monte Carlo model of epitaxial graphene growth}}
\author{
\textbf{Bartomeu Monserrat S\'{a}nchez}\\
\vspace{1.0cm}
Imperial College London\\
Project code: CMTH--Vvedensky--3\\
Supervisor: D. D. Vvedensky\\
Assessor: W. M. C. Foulkes
\date{May 3, 2011}
}
\maketitle

\newpage
\thispagestyle{empty}
\mbox{}

\newpage

\section*{Acknowledgments}

\thispagestyle{empty}
\mbox{}

The research described in this thesis corresponds to the final year Physics MSci project. I was mentored by Dimitri Vvedensky and benefited from his knowledge continuously. I am really fortunate to have worked with such an inspiring supervisor.  

I collaborated closely with Jonathan Lloyd-Williams, who contributed enormously to this work. Many interesting discussions allowed us to progress steadily. Each of us kept their own code capable of producing all the data presented in this thesis. However, we split some of the analysis and I thank Jonathan for providing the data for the $j = 2$ and rate equations comparison plots. 

Mireia Crisp\'{i}n Ortuzar, Raphael Houdmont, Josep Monserrat and Kevin Troyano read the manuscript and their comments and suggestions improved it greatly. I am really thankful to all of them. Also, they all made my life in London interesting, amusing and fun, together with Freddie, Daniel, Ryan, Bhavika, Grace, Helena, Virginia and many others. 

Mireia has been my inspiration and has pushed me forward with her love in every enterprise I have taken. Finally, I would like to thank my parents for giving me the opportunity to study at Imperial and always supporting me in my studies.

\newpage
\thispagestyle{empty}
\mbox{}

\begin{abstract}
In this thesis we present a kinetic Monte Carlo model for the description of epitaxial graphene growth. Experimental results suggest a growth mechanism by which clusters of 5 carbon atoms are an intermediate species necessary for nucleation and island growth. This model is proposed by experimentally studying the velocity of growth of islands which is a highly nonlinear function of adatom concentration. In our simulation we incorporate this intermediate species and show that it can explain all other experimental observations: the temperature dependence of the adatom nucleation density, the equilibrium adatom density and the temperature dependence of the equilibrium island density. All these processes are described only by the kinematics of the system. 
\newpage
\thispagestyle{empty}
\mbox{}

\end{abstract}

\pagenumbering{roman}
\setcounter{page}{1}
\tableofcontents

\chapter{Introduction}

\pagenumbering{arabic}
\setcounter{page}{1}

\section{Scope of this thesis}

Graphene is a 2-dimensional layer of graphite, formed by carbon atoms arranged in an hexagonal lattice. For a long time, graphene was used as the starting point for the theoretical study of carbon allotropes, such as carbon nanotubes, which can be described by the rolling of a graphene sheet into the form of a cylinder. But it was thought that graphene was merely a theoretical construct because thermal fluctuations would render the isolated 2-dimensional sheets unstable. Nonetheless, it was reported by Novoselov and co-workers in 2004 \cite{ref:firstreport} that graphene sheets had been isolated and some of their physical properties measured. Since then, it has attracted the attention of the physics and materials communities \cite{ref:rise.graphene} due to its interesting physical properties and its potential for technological applications. The 2010 Physics Nobel Prize was awarded to Andre Geim and Konstantin Novoselov

\begin{quote}
\textit{for groundbreaking experiments regarding the two-dimensional material graphene}
\end{quote}
as described by the awarding body.

Recent efforts have focused on synthesising graphene sheets by means of epitaxial growth, one of the most promising routes to large scale and high quality graphene production. There exist ample studies characterising the properties of graphene sheets grown on a variety of substrates. However, very little work has been done on the growth kinetics of the graphene sheets. The first experimental studies only became available in 2008 \cite{ref:loginova2008,ref:loginova2009}. The remarkable behaviour observed in these studies is very different from that of known metal-on-metal growth systems. Initial theoretical studies inspired by the experimental findings have been conducted using a rate equations approach \cite{ref:zangwill}, and these studies have been able to describe the main features of the experimental results. 

In this thesis we present a kinetic Monte Carlo model for theoretical studies of epitaxial graphene growth kinetics. The model is based on the basic principles found experimentally and used in the rate equations study, and provides further insight into the physical processes governing the system.

\section{Thesis outline}

Chapter \ref{ch:epitaxialgraphene} is organised in two parts, and it describes some background information relevant to the project. First we describe the physical properties and technological potential of graphene, which have made it an attractive system for both the physics and materials communities. In the second part we present epitaxial growth, discussing the physical processes involved together with the experimental techniques used in the field.

In Chapter \ref{ch:previous} we give a detailed overview of the current understanding of epitaxial graphene growth. We survey the most relevant experimental and theoretical results describing the kinetics of epitaxial graphene growth, and abstract the principal conclusions to be incorporated in a model to study the system further.

In Chapter \ref{ch:kmc} we introduce in detail the kinetic Monte Carlo (KMC) technique, focusing on the N-fold way algorithm. We also provide motivation for the use of this method in the study of epitaxial growth systems. 

In Chapter \ref{ch:standard} we present a simple standard KMC model for epitaxial growth, exemplifying the N-fold way algorithm. This simulation will be used as a starting point for more complex models appropriate for the study of the graphene system. 

In Chapter \ref{ch:tetramer} we present a KMC model that includes the main features observed in graphene growth experiments. All the experimental observations are explained physically within this model, identifying the most important processes governing the kinetics of epitaxial graphene growth.

In Chapter \ref{ch:conclusions} we summarise the most relevant results and discuss lines of research that can be followed from the present work.

\chapter{Graphene and epitaxial growth} \label{ch:epitaxialgraphene}

In this chapter we first present the most relevant physical properties of graphene and discuss its technological potential. We then look at both theoretical foundations and experimental techniques in epitaxial growth, which is most probably the method of choice for large scale and high quality graphene production.

%I THINK I SHOULD KEEP THESE SECTIONS SHORT

\section{Graphene}
\subsection{Physical properties}
The physical properties of graphene were recently reviewed in Ref.\cite{ref:graphenereview}. Graphene is a 2-dimensional layer formed by sp$^2$-hybridised carbon atoms arranged in a hexagonal lattice and separated by $a \simeq$ 1.42 \AA. The hybrid orbitals give rise to $\sigma$ bonds between atoms. The remaining orbital, $p_z$, is perpendicular to the planar structure and forms covalent bonds between neighbouring carbon atoms, leading to a half filled $\pi$ band. This band can be described using a tight-binding approach with a single hopping matrix element between neighbouring atoms $-t$. The resulting band structure, first calculated by Wallace in 1947 \cite{ref:original1947}, has a dispersion relation 

\begin{equation}
%E_{\pm}(\textbf{q})\approx\pm v_F |\textbf{q}| + \mathcal{O}[(q/k_F)^2] \label{eq:dispersion}  
E(\textbf{k}) = \pm t\sqrt{3+2\cos(\sqrt{3}k_y a)+4\cos\left(\frac{\sqrt{3}}{2}k_y a\right)\cos\left(\frac{3}{2}k_x a\right)} \label{eq:dispersion}
\end{equation}
for wavevector ($k_x$,$k_y$). 
\begin{figure}[h!]
\centering
\includegraphics[scale=0.3]{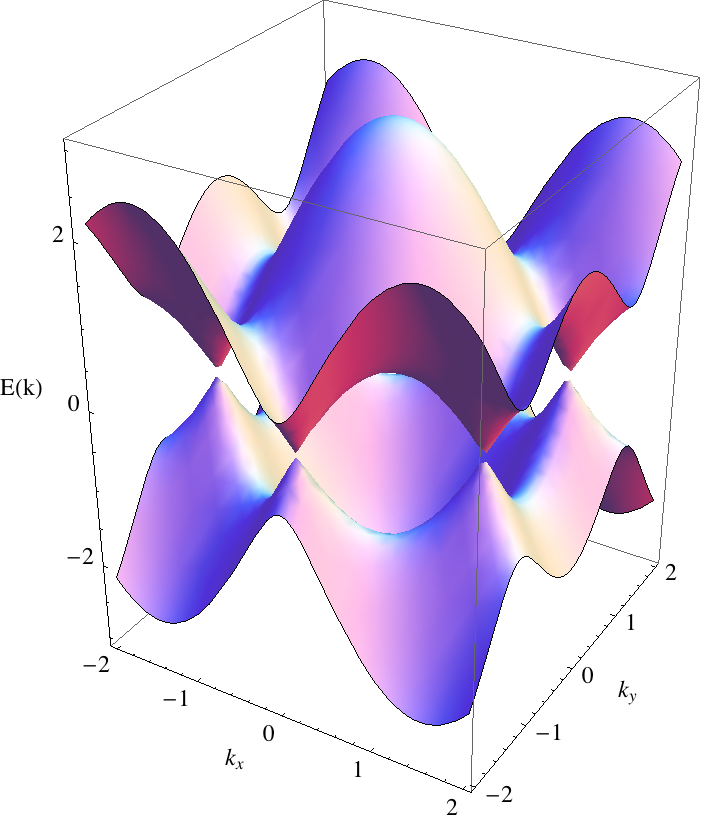}
\caption{\label{fig:dispersion}Electronic dispersion relation of graphene obtained with a tight-binding calculation with nearest neighbour interactions, given by Eq.(\ref{eq:dispersion}). The energy $E(\textbf{k})$ is in units of $t$.}
\end{figure}

Graphene is a zero gap semiconductor, where the Fermi surface consists only of six Fermi points at the edge of the Brillouin zone, as shown in Fig. \ref{fig:dispersion}. Expanding about the Fermi points $\textbf{k}_F=\textbf{k}-\textbf{q}$ in reciprocal space we find $E_{\pm}(\textbf{q})\approx\pm v_F |\textbf{q}| + \mathcal{O}[(q/k_F)^2]$ for constant $v_F\simeq$ 10$^6$ ms$^{-1}$. This parallels the dispersion relation of ultrarelativistic particles described by the massless Dirac equation, so the Fermi points are commonly called Dirac points. It is this dispersion that determines most of the singular physical properties of graphene and it means that graphene is a laboratory condensed matter system to test (2+1)--dimensional quantum electrodynamics \cite{ref:qed}.

As reported in Ref.\cite{ref:novoselov.nature} the experimental study of some of the physical properties confirmed the existence of massless Dirac carriers in graphene. For instance, their cyclotron mass depends on the square root of the density of states, and the integer quantum Hall effect occurs at half-integer filling factors, both characteristic of massless Dirac fermion systems.

\subsection{Technological applications}
The physical properties of graphene result in large carrier mobilities that persist at room temperature, even with the presence of doping species. This means graphene holds the potential to become a replacement for silicon in the electronics industry \cite{ref:rise.graphene}, along with a wider variety of applications.

Transistors and diodes need the presence of a band-gap for their operation, %\cite{ref:switched.on}
so standard graphene sheets are not appropriate. Researchers have explored alternative graphene-based structures with the presence of a band-gap, for instance graphene nanoribbons \cite{ref:ribbons.theo,ref:ribbons.exp} where the band-gap is proportional to the ribbon width, or graphene nanomeshes \cite{ref:mesh} where the graphene sheets are punched with an array of nanoscale holes. Graphene-made transistors in the GHz scale were recently reported by IBM researchers \cite{ref:tech.ibm,ref:tech.ibm2,ref:wu2011} with performances superior to those of similar silicon transistors.

Other examples are studies on the large heat conductivity of graphene \cite{ref:tech.heat,ref:tech.heat1} with potential applications in nanoelectronics where large heat dissipation is needed.

\section{Epitaxial growth}

Epitaxial growth is the name given to the process of producing epitaxial thin films on substrates. It is a widespread technique with applications ranging from the production of semiconducting devices to nanotechnology and it has also attracted the scientific community because of the complex atomic processes involved.

\subsection{Theoretical description}

Epitaxial growth can be classified in three so-called modes, first introduced in the seminal work by Bauer \cite{ref:modes}. Following Ref.\cite{ref:venables.book},
%, ref:surface.book, ref:venables.review, ref:venables.review.modern}, 
the different growth modes can be understood by thermodynamic arguments. In the \textit{layer}, or Frank--van der Merwe mode, the atoms are more strongly attracted to the substrate than to themselves, and the epitaxial film is formed layer after layer. Quantitatively, it arises in the deposition of material A on B when the surface free energies $\gamma_i$ for $i=A,B$ obey 

\begin{equation}
\gamma_A + \gamma_{int} < \gamma_B \label{eq:free.energies}
\end{equation}
for interface free energy $\gamma_{int}$ between surfaces A and B. This follows because the free energy needs to be minimised at equilibrium, and the inequality requires to maximise the area covered by deposit A. The \textit{island}, or Volmer--Weber mode, results when the atoms are more strongly attracted by each other than to the substrate, and multilayer (3-dimensional) islands form. Quantitatively, $\gamma_A + \gamma_{int} > \gamma_B$. A third hybrid mode termed Stranski--Krastanov growth mode, or \textit{layer-plus-island}, arises because the interface energy $\gamma_{int}$ increases as the layer thickness increases, so island growth starts as a layer mode but turns into an island mode. 

In the present work we only consider submonolayer growth when the first layer is forming, because we only reach fractional coverages of the lattice and the system is purely 2-dimensional.

In the atomic regime there are many processes during deposition and growth. Atoms are deposited at a certain rate on the substrate, and then undergo a series of processes. They can re-evaporate, diffuse over the substrate or along island edges, nucleate to form islands, join growing islands or detach from islands. Temperature is a key parameter because the different atomic processes are thermally activated. In the present work we incorporate these atomistic processes into a kinetic Monte Carlo model.

\subsection{Experimental techniques}

Molecular beam epitaxy (MBE) \cite{ref:mbe} is the most widespread technique used in epitaxial growth. A beam of atoms or molecules is deposited on a previously prepared substrate kept at high temperatures to allow the arriving particles to diffuse over its surface. The deposition is carried out under ultra-high vacuum conditions in order to minimise impurities. Control over the beam allows films grown using MBE to be of very high quality and to have the desired properties. Typical deposition rates are $\sim$1 ML/s, which are high enough to significantly reduce the incorporation of impurities into the growing material. Another technique used in epitaxial growth is vapour phase epitaxy where the substrate is placed in contact with a gas containing the deposit elements, and reactions between the two lead to epitaxial growth.

The grown sheets can be observed with high accuracy using different microscopy techniques such as scanning tunneling microscopy or atomic force microscopy. Furthermore, the growing process can be monitored using low energy electron microscopy (LEEM). For example, in some experiments \cite{ref:loginova2008,ref:loginova2009} relevant for the work described in this thesis the changes in reflectivity of a LEEM are used to infer the carbon adatom concentration on the substrate.

\chapter{Experimental and theoretical results on epitaxial graphene growth} \label{ch:previous}

Epitaxial growth is one of the most promising techniques currently being explored to synthesise high quality graphene sheets with the appropriate properties for technological applications. Much ongoing research concentrates in the production of such epitaxial sheets and their characterisation for a variety of substrates with different physical properties have been reported. However, studies of the growth kinetics of graphene are just beginning both experimentally and theoretically and here we present a review of the current status of such studies.

\section{Experimental results} \label{subsec:exp}

In this section we present the most relevant experimental results that motivated the present work. They are the first that look at the kinetics of the growth of epitaxial graphene.

\subsection{Kinetics of epitaxial graphene growth}

The most relevant experimental observations of the kinetics of epitaxial graphene growth have been reported by Loginova and co-workers in Refs.\cite{ref:loginova2008,ref:loginova2009}. They deposit pure carbon atoms and ethylene molecules on ruthenium Ru(0001) and iridium Ir(111) substrates under ultra-high vaccum conditions and high substrate temperatures. Low Energy Electron Microscopy (LEEM) is used to monitor the growth of graphene islands and the local carbon adatom concentration on the substrate.

\begin{figure}[h!]
\centering
\includegraphics[scale=0.5]{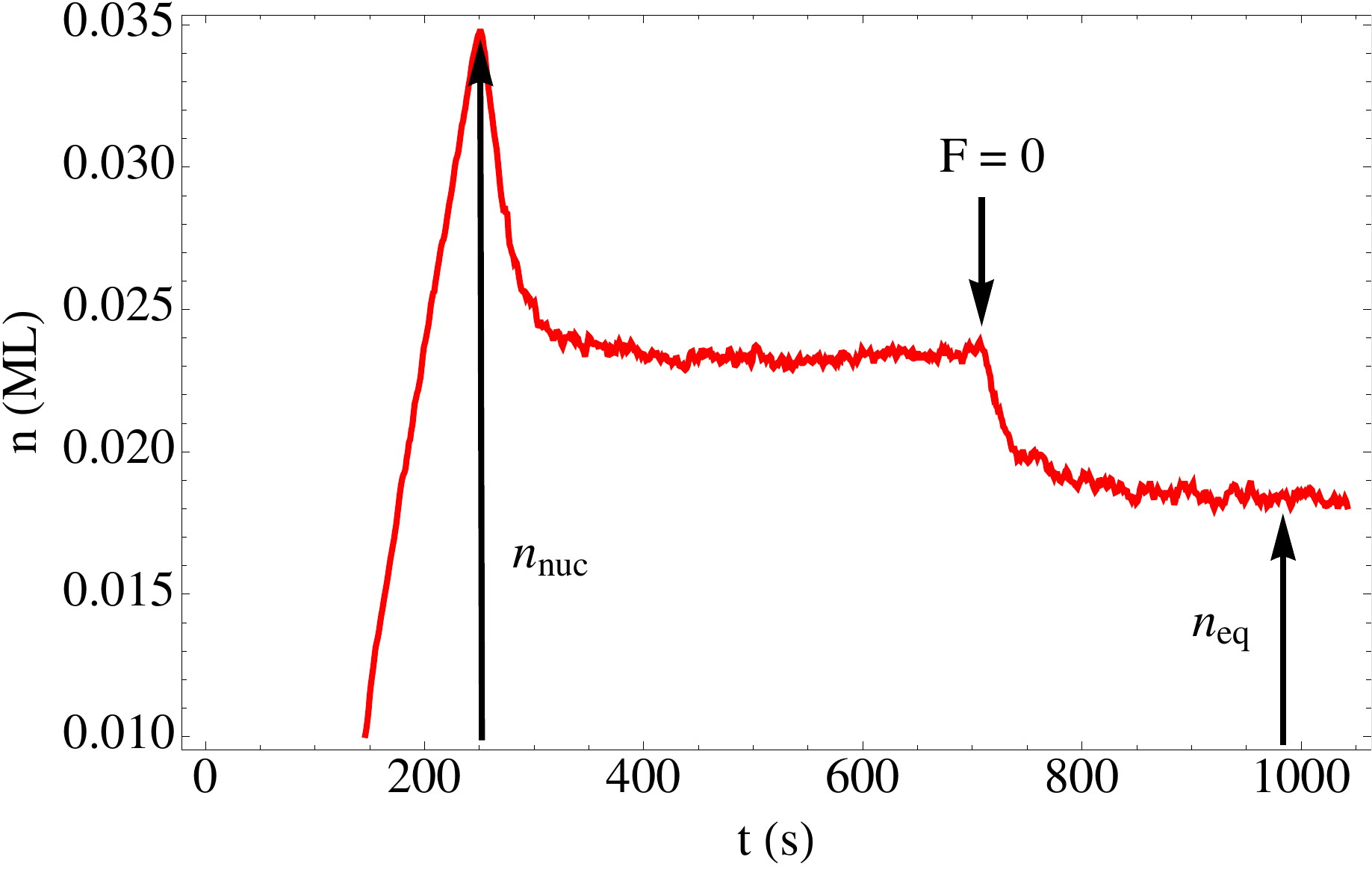}
\caption{\label{fig:adatom}Typical adatom density $n$ curve as a function of time found in epitaxial growth experiments, adapted from Loginova \textit{et al.} \cite{ref:loginova2008}.}
\end{figure}

A typical adatom concentration profile that will appear throughout this work is as shown in Fig. \ref{fig:adatom}. A deposition flux $F$ is turned on at about time $t = $ 150 s. Initially the adatom concentration increases almost linearly with time due to the constant flux, in the example given of $F =$ 0.0023 ML/s, small for typical epitaxial experiments. At about $t =$ 250 s the adatom density curve reaches a maximum, called $n_{nuc}$ because it approximately corresponds to the onset of nucleation.  Islands start nucleating and adatoms start disappearing from the substrate because they attach to islands. This decrease continues for some 100 s until an equilibrium between islands and adatoms is reached. Later on, at $t = $ 700 s, the deposition flux is turned off, quickly leading to a different equilibrium concentration between adatoms and islands, smaller than the previous equilibrium because adatoms are no longer being incorporated externally into the system. The adatom density in this regime is labelled $n_{eq}$.

Note the definition of $n_{nuc}$ can be slightly confusing in systems where the critical island size is not well-defined. The critical island size $i$ is an island size above which adatom attachment can be considered as irreversible, so that nucleation can be defined as the process of an island going from size $i \rightarrow (i+1)$. Then, in systems with no critical island size, large clusters that eventually lead to islands can start appearing before the maximum of the adatom density profile is reached. Throughout this work we are going to define $n_{nuc}$ as the maximum of the adatom density profile.

\begin{figure}[h!]
\centering
\includegraphics[scale=0.4]{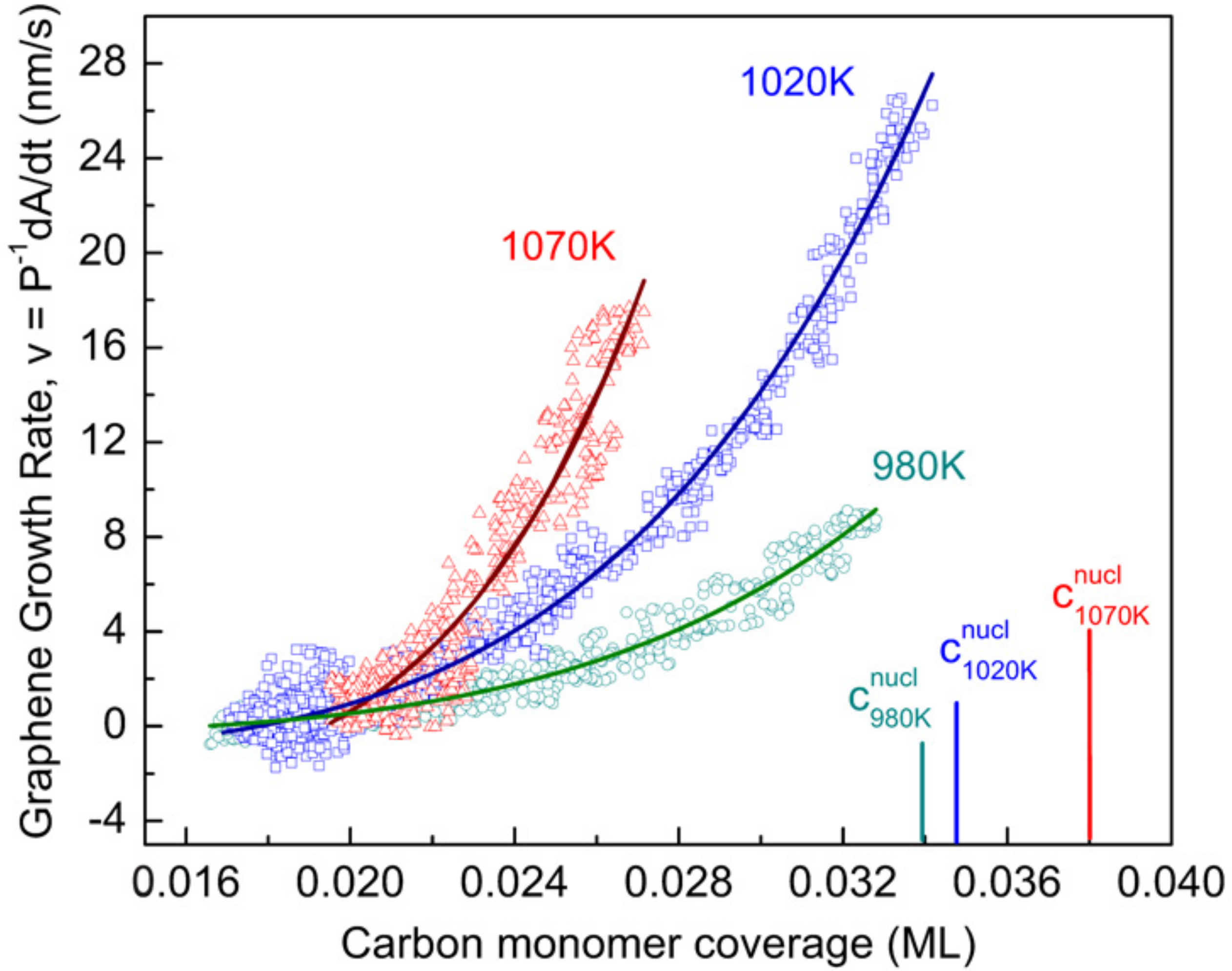}
\caption{\label{fig:nonlinearexp}Experimental observation of nonlinear island growth velocity as a function of adatom concentration. In the plot, $c^{nucl}$ corresponds to $n_{nuc}$ in the text. Taken from Loginova \textit{et al.} \cite{ref:loginova2008}.}
\end{figure}

In their studies, Loginova and co-workers find that the island growth velocity presents a nonlinear dependence on the carbon adatom concentration as shown in Fig. \ref{fig:nonlinearexp}. The island growth velocity $v$ is defined as $v = P^{-1}dA/dt$ for island perimeter $P$ and area $A$. In most growth systems the island growth velocity is found to be proportional to the supersaturation of adatoms, i.e. proportional to the difference in adatom concentration $n$ and adatom equilibrium concentration $n_{eq}$, $v = C(n-n_{eq})$ for constant $C$. To explain the nonlinear relationship found for graphene growth, Loginova and co-workers propose a model in which the energy barrier for monomer attachment to islands is larger than the barrier for the formation of clusters of $m$ carbon atoms and their posterior attachment to islands. They assume that the growth velocity is proportional to the supersaturation of clusters rather than adatoms. Following Ref.\cite{ref:loginova2008}, the concentration of $m$-clusters $c^{(m)}$ in a supersaturated adatom sea has an exponential dependence on the energy difference between $m$ isolated carbon atoms and the energy needed to form an $m-$atom cluster $E_m$,

\begin{equation}
c^{(m)} = e^{(m\mu - E_m)/k_BT} = \left(\frac{n}{n_{eq}}\right)^me^{-E_m/k_BT},
\end{equation}  
where the sea of carbon adatoms is assumed to be an ideal lattice gas with carbon chemical potential $\mu = k_BT\ln(n/n_{eq})$. The island growth velocity as a function of adatom concentration is then

\begin{equation}
v = C_m(c^{(m)}-c^{(m)}_{eq}) = B\left[\left(\frac{n}{n_{eq}}\right)^m-1\right]
\end{equation}
where $B = C_me^{-E_m/k_BT}$ for $C_m$ the proportionality constant in the velocity dependence in cluster supersaturation. Fitting the data in Fig. \ref{fig:nonlinearexp}, Loginova and co-workers find a best estimate of $m\simeq$ 5. This means that in their model clusters formed by 5 carbon atoms are an intermediate species that determines the growth kinetics of epitaxial graphene growth. This intermediate species could be formed on the graphene free lattice and diffuse over the substrate to attach to islands as is assumed throughout Ref.\cite{ref:loginova2008}. This model is the one we are going to take in our KMC approach to the description of the system. However, it is pointed out by Loginova and co-workers that the 5-clusters could instead form only near island edges at the moment of attachment to graphene.

\begin{figure}[h!]
\centering
\includegraphics[scale=0.4]{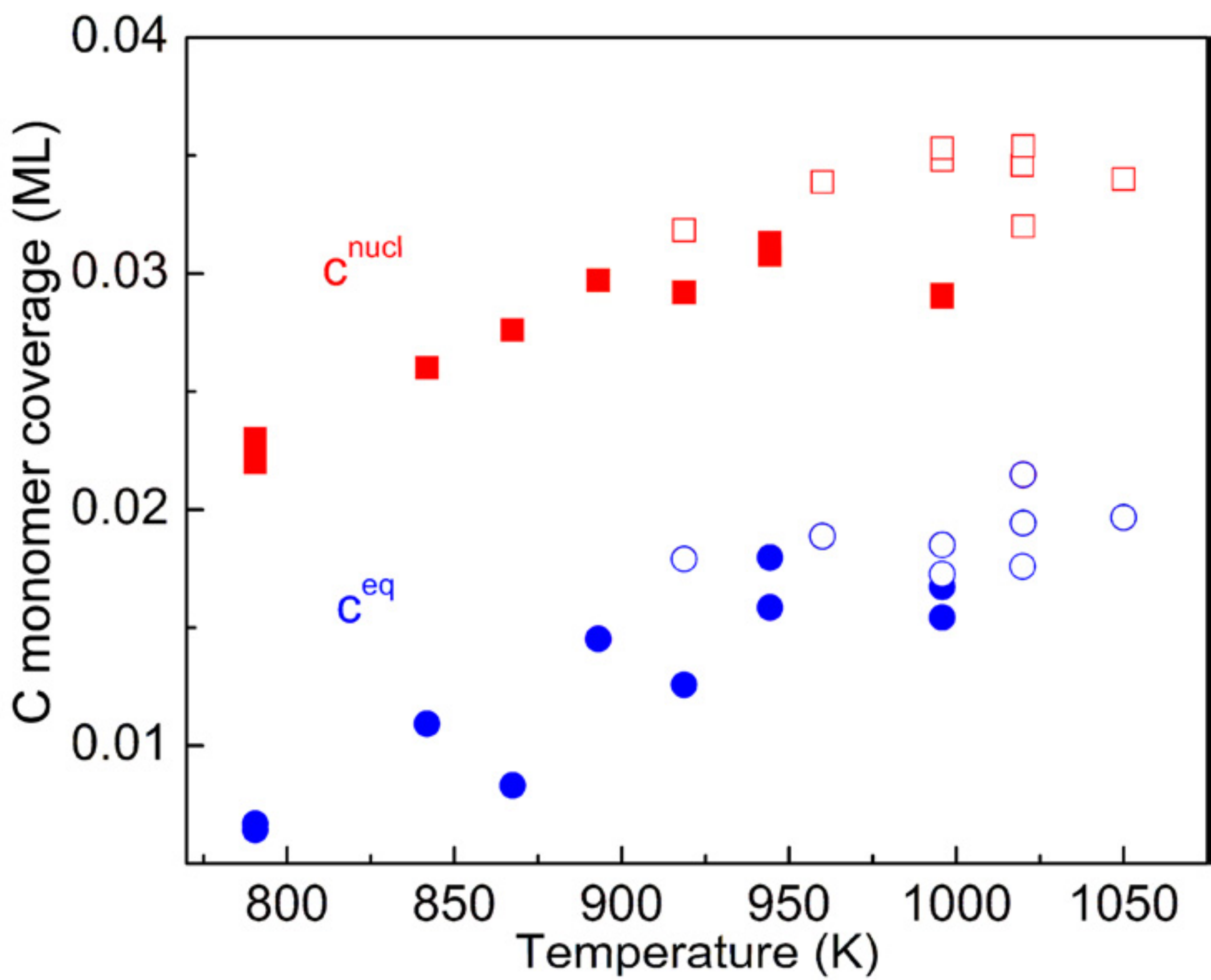}
\caption{\label{fig:nucexp}Adatom density at nucleation $n_{nuc}$ ($c^{nucl}$ in the plot) and at equilibrium $n_{eq}$ ($c^{eq}$ in the plot) as a function of temperature. The data corresponds to deposition of carbon (filled labels) or ethylene (emptly labels). Taken from Loginova \textit{et al.} \cite{ref:loginova2009}.}
\end{figure}

Further observations in the graphene growth system that differ from other known growth scenarios are the temperature dependence of adatom density at nucleation and the adatom density at equilibrium. 

It can be seen in Fig. \ref{fig:nucexp} that the nucleation density increases with increasing temperature. Most growth systems are diffusion limited, so that increasing the temperature results in higher adatom mobility and earlier nucleation. Therefore, the nucleation density usually decreases with increasing temperature unlike for graphene. 

It can also be observed in Fig. \ref{fig:nucexp} that the nucleation concentration is roughly twice as large as the equilibrium density $n_{nuc}\sim2n_{eq}$. This indicates a small energy barrier to adatom detachment from graphene sheets and a correspondingly large attachment barrier for adatoms to growing graphene. This observation is in agreement with the model by which the dominant species in the kinetics of graphene growth is an intermediate 5-cluster.

Another experimental observation that remains unpublished is the behaviour of the island density at equilibrium. In agreement with known growth systems (described below in Chapter \ref{ch:standard}) the island density decreases with increasing temperature. However, the decrease in the graphene system is found to be much larger than in other systems, suggesting the existence of a mechanism that allows the occurence of a large number of nucleations at low temperatures. We thank Elena (Loginova) Starodub for providing the data related to island density.

We finally note that the growth mechanism of graphene is different for other metal substrates. For instance, for Ru(0001) and Ir(111), the deposition of molecules containing hydrogen and carbon decompose rapidly and growth is determined by individual carbon atoms. However, theoretical and experimental studies \cite{ref:zhang2011,ref:treier2010} indicate that graphene growth on copper is determined by the deposited molecules, and that only at late stages of the graphene formation process is hydrogen released from the composites.

\subsection{Graphene nanoclusters}

Another set of experiments \cite{ref:cui2011,PhysRevLett.103.166101} looks more carefully at the very initial steps of graphene growth when nucleation occurs. The studies are both on Ru(0001) and Ir(111) as the experiments by Loginova and co-workers reported above. Small carbon nanoclusters of sizes of the order of tens of carbon atoms are observed on both substrates. They are described as a possible predecessor for graphene islands, and could be understood as a critical island size for graphene. 

\begin{figure}[h!]
\centering
\includegraphics[scale=0.4]{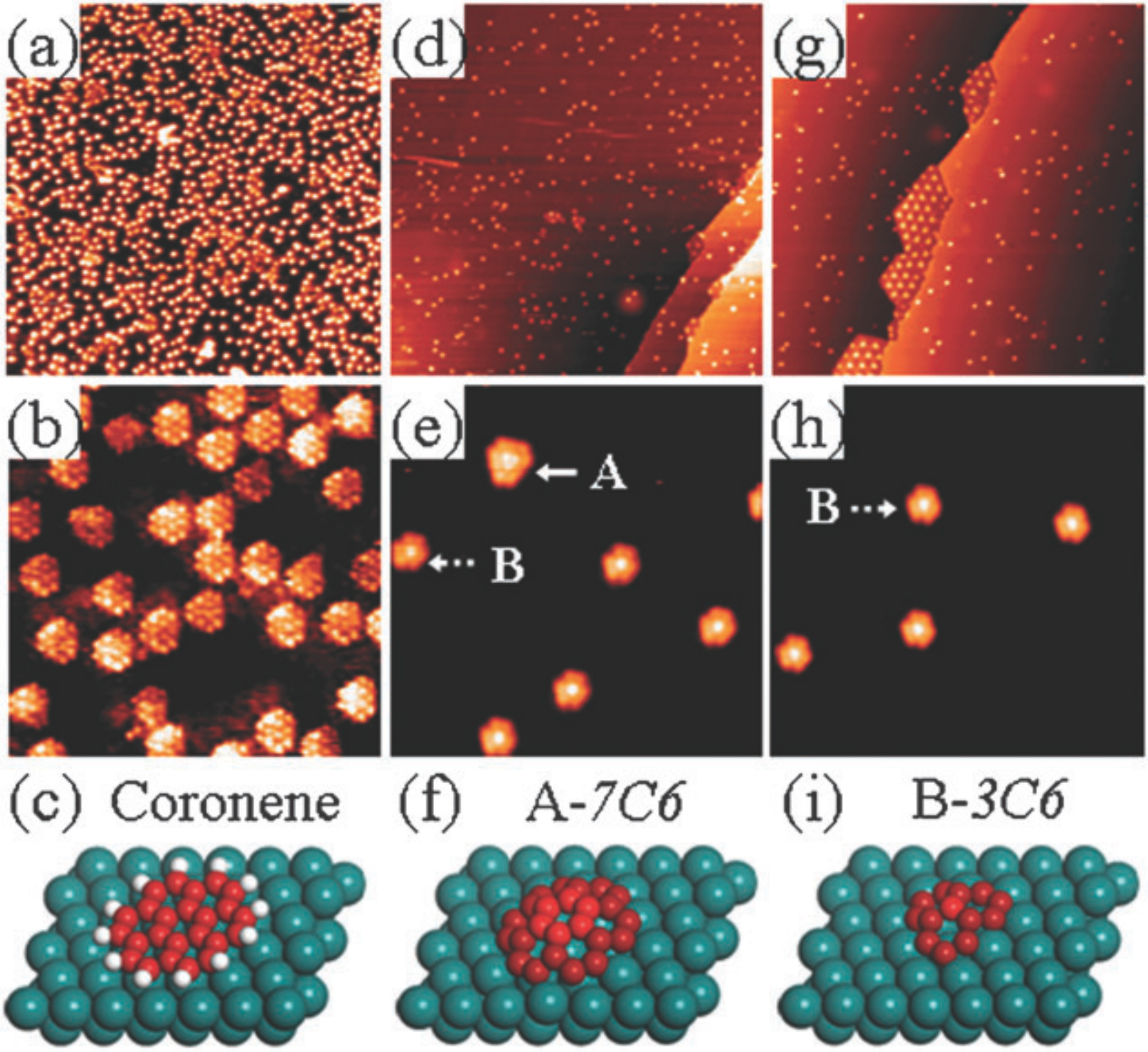}
\caption{\label{fig:dome}Images of coronene and dome-like carbon nanoclusters on Ru(0001), together with a depiction of the structure of the clusters. Taken from Cui \textit{et al.} \cite{ref:cui2011}.}  
\end{figure}

In both cases, the observed nanoclusters have a dome-like shape where the adatoms at the perimeter of the quasi-circular islands are strongly attached to the substrate and the adatoms in the center are highly detached from it. This structure can be seen in Fig. \ref{fig:dome}.

Other studies of epitaxial graphene growth on Rhodium Rh(111) reported in Ref.\cite{doi:10.1021/nl103053t} also indicate the presence of these nanoclusters. The study indicates that the nanoclusters are mobile on Rh(111), so they would not correspond to immobile critical islands. However, the experiment is performed on a different substrate to the above experiments, and the experimental method differs as well. This means that the results might not be directly comparable to the above.

\section{Theoretical results} \label{subsec:theo}

Based on the experiments reported by Loginova and co-workers in Refs.\cite{ref:loginova2008,ref:loginova2009}, Zangwill and Vvedensky proposed a rate equations (RE) model \cite{ref:zangwill} to describe epitaxial graphene growth. RE are described in the next chapter. They incorporate the intermediate 5-clusters and propose a system evolution determined by a set of coupled differential equations for the adatom density $n$, the 5-atom cluster density $c$ and the island density $N$ as follows

\begin{eqnarray}
\frac{dn}{dt} &=& F - iDn^i + iKc - DnN + K'N,\\
\frac{dc}{dt} &=& Dn^i - Kc - D'cN - jD'c^j,\\
\frac{dN}{dt} &=& D'c^j, 
\end{eqnarray}
The parameters are carbon atom deposition flux $F$, adatom diffusion rate $D$, cluster diffusion rate $D'$, cluster dissolution rate $K$ and adatom detachment rate from islands $K'$, all assumed to be of the Arrhenius form because they are thermally activated. The index $i =$ 5 is the cluster size, and the index $j$ is an unknown for the number of clusters that need to come together to form an island. %The initial problem has 5 free parameters, but the experimental results on the adatom density at the equilibrium with non-zero flux and $n_{eq}$ put constraints on the system reducing the problem to a 3 parameter fit.

\begin{figure}[h!]
\centering
\includegraphics[scale=0.48]{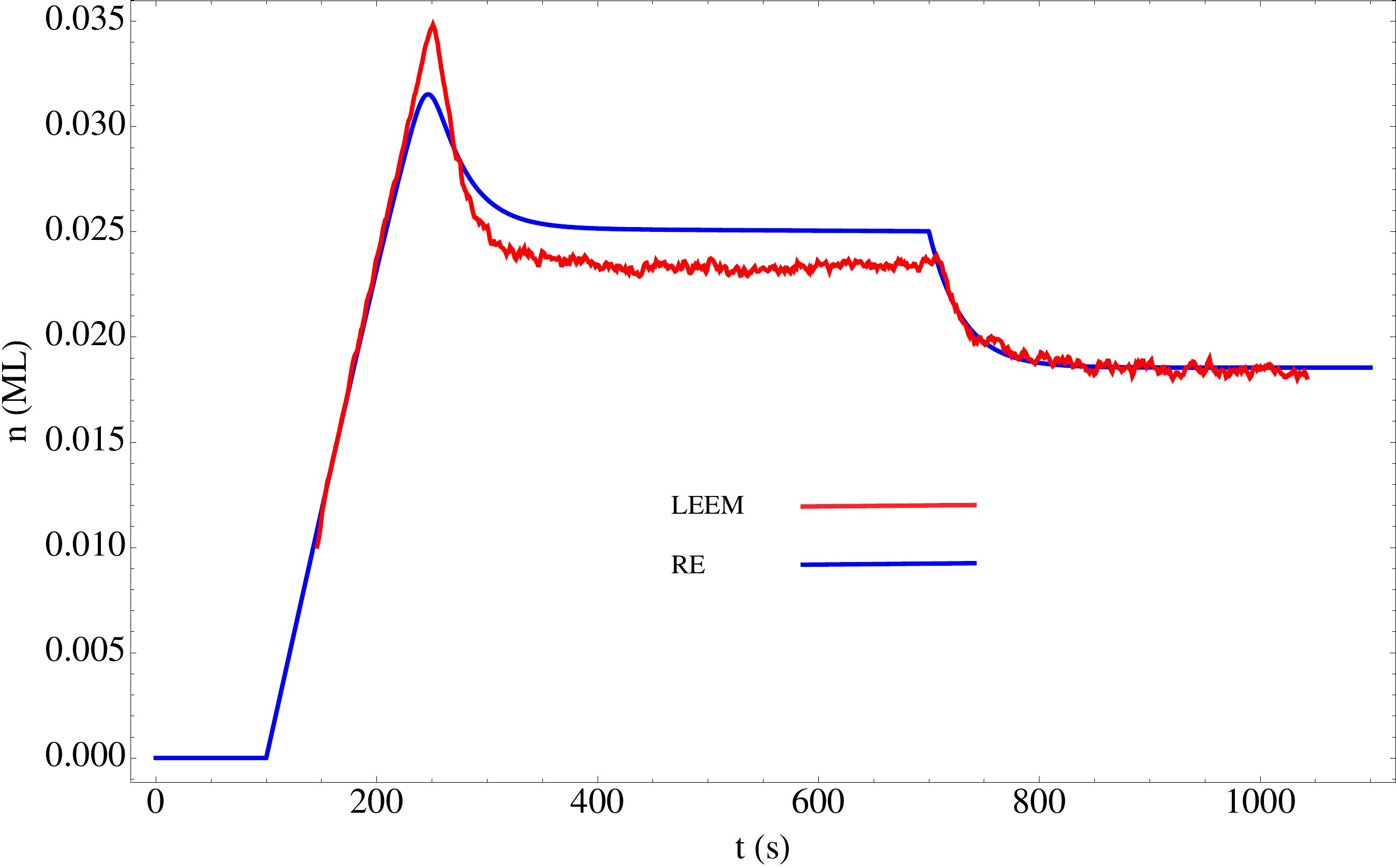}
\caption{\label{fig:refit}Rate equations comparison with LEEM data from Ref.\cite{ref:loginova2009} for the adatom density profile. Adapted from Zangwill and Vvedensky \cite{ref:zangwill}.}
\end{figure}

The above set of equations together with an appropriate choice of values for the parameters reproduces the main features of the adatom density curve in Ref.\cite{ref:loginova2009} as shown in Fig. \ref{fig:refit}. The temperature dependence of the adatom density at nucleation is well-reproduced within the RE by introducing the variable $j$ and taking a value $j\geq$ 6, and with the presence of the cluster dissociation rate. However, the temperature dependence of island density at equilibrium cannot be explained with the above RE model. Discrepancies with the experimental data are attributed to spatial effects that are not included in the RE approach.

\section{Kinetic Monte Carlo model}

Based on the above experimental and theoretical results of graphene growth, a KMC model to describe epitaxial growth of graphene sheets should incorporate the following basic ingredients: three different species \cite{ref:loginova2008,ref:loginova2009} corresponding to the experimentally observed carbon monomers, carbon 5-clusters, and graphene sheets; and a large critical nucleus size \cite{ref:cui2011,PhysRevLett.103.166101}.

These basic components of the model should form the basis for the description of the experimental observations on the kinetics of epitaxial graphene growth \cite{ref:loginova2008,ref:loginova2009}:

\begin{enumerate}
\item The adatom density at the onset of nucleation increases with increasing temperature.
\item The adatom density at equibrium $n_{eq}$ is roughly half of the adatom density at the onset of nucleation $n_{nuc}$, namely, $n_{nuc} \sim 2n_{eq}$.
\item The island density at equilibrium decreases rapidly with increasing temperature.
\end{enumerate}

In the rest of this thesis we will present such a KMC model and describe the relevant physical processes in epitaxial graphene growth within the model.

\chapter{Kinetic Monte Carlo} \label{ch:kmc}

In this chapter we present the kinetic Monte Carlo (KMC) method as used in the study of epitaxial growth. The most widespread algorithm used in the field is the so-called \textit{N-fold way} algorithm, which is adopted in this thesis.

\section{Theoretical studies and computer simulations}

%Computer simulations have a side-by-side status with experimental science. Theories can be tested using a computer, and system conditions unattainable in the laboratory can be reproduced. Furthermore, parameters can be changed individually giving a greater control over the study of particular processes. 
In the field of many-body dynamics simulations, and in particular epitaxial growth systems, there are various approaches with different levels of approximation and capabilities.

A completely deterministic mean-field approach called rate equations (RE) is commonly used in studies of epitaxial growth. RE are a finite set of coupled first order differential equations for the densities of the different species in an epitaxial system and with a rate associated with each possible process. The strength of the technique is its simplicity, but the pay-off is in accuracy. Standard RE cannot include any spatial information, but more complex approaches using capture numbers can encapsulate it. Zangwill and Vvedensky have used RE to describe the epitaxial graphene growth system \cite{ref:zangwill} as described above in Section \ref{subsec:theo}.  

Molecular dynamics (MD) simulations are used for detailed analysis of the dynamics of a system. The forces between different components are calculated and the system is evolved according to these. Both classical and quantum approaches to MD simulations are used, and major advances in the field, for instance the Car-Parrinello method \cite{ref:car.parrinello} incorporating density functional theory into MD, have led to widespread use of the method. However, the timescales and system sizes that can be explored remain small, restricting the applicability of the technique.

KMC looks at intermediate time and length scales and has proved appropriate for the description of epitaxial growth systems.

\section{Kinetic Monte Carlo}

In this section we describe the KMC technique. We first present the N-fold way algorithm used in this thesis and next we discuss why the KMC method can be used in the description of epitaxial growth processes.

\subsection{N-fold way algorithm}

The KMC method is based on the use of random numbers to simulate the time evolution of a system in discrete steps, in such a way that each step represents a move of the entire system from one state to another. We are going to concentrate on the N-fold way algorithm, first introduced by Bortz, Kalos and Lebowitz in 1975 \cite{ref:bortz1975}. 

Following Ref.\cite{ref:maksym1988}, a rate $r_i$ is associated with process $i$, and rates are usually taken to be of the Arrhenius form because they are thermally activated,

\begin{equation}
r_i = \nu_0 e^{-E_i/k_BT}
\end{equation}
where $\nu_0 = 2k_BT/h$ is of the order of the atomic vibrational frequency, and $E_i$ is the energy barrier associated with process $i$. $T$ is the temperature, $k_B$ is Boltzmann's constant and $h$ is Planck's constant. This form for the rates will imply in a KMC model that events that are more likely to occur will happen more frequently. At each time step, all rates are calculated for the given configuration of the system.

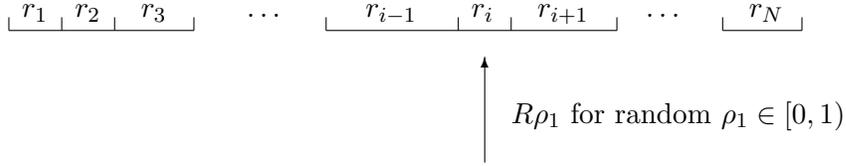
\begin{figure}

\centering

\begin{picture}(0,0) 
\put(-150,0){\line(1,0){70}}
\put(-30,0){\line(1,0){110}}
\put(120,0){\line(1,0){30}}
\put(-150,0){\line(0,1){5}}
\put(-130,0){\line(0,1){5}}
\put(-110,0){\line(0,1){5}}
\put(-80,0){\line(0,1){5}}
\put(-30,0){\line(0,1){5}}
\put(20,0){\line(0,1){5}}
\put(40,0){\line(0,1){5}}
\put(80,0){\line(0,1){5}}
\put(120,0){\line(0,1){5}}
\put(150,0){\line(0,1){5}}

\put(30,-50){\vector(0,1){40}}

\put(-145,5){$r_1$}
\put(-125,5){$r_2$}
\put(-100,5){$r_3$}
\put(-60,5){$\ldots$}
\put(-15,5){$r_{i-1}$}
\put(25,5){$r_i$}
\put(50,5){$r_{i+1}$}
\put(91,5){$\ldots$}
\put(130,5){$r_N$}

\put(40,-35){$R\rho_1 \mbox{ for random } \rho_1\in[0,1)$}

\end{picture}

\vspace{2cm}

\caption{\label{fig:nfoldway}Schematic of the N-fold way algorithm selecting the rate $r_i$ to occur in the given time step. The size of the boxes containing the rates is proportional to the rates themselves.}

\end{figure}

To select the process that is going to occur at a given time step, a total transition rate $R$ is constructed as 

\begin{equation}
R = \sum_{i = 1}^N r_i
\end{equation}
for a total of $N$ possible processes in a given configuration of the system. A number $\rho_1\in[0,1)$ uniformly distributed is calculated, and then the process $j$ such that

\begin{equation}
\sum_{i=1}^{j-1} r_i \leq R\rho_1 <\sum_{i=1}^j r_i
\end{equation}
is selected to take place in the given time step. This is shown in Fig. \ref{fig:nfoldway}. 

Practically, possible events can be grouped together and the search time for the event is reduced. For instance, all free adatoms will have equal probability of diffusing into any of the four nearest-neighbour sites in a square lattice simulation, and all these events can be put into a subgroup. Constructing these subgroups reduces the total rate to a set of partial sums, and the search then proceeds first by identifying the relevant subgroup and then searching only within this subgroup. This search method is referred to as the \textit{binning method} \cite{ref:maksym1988} or 2-level scheme as the search is done in two levels. There are methods available \cite{PhysRevE.51.R867} which are up to 7 times faster than the binning method and are based on a further subdivision of the list of possible events, refered to as $K$-level schemes for $K$ subdivisions. %In this work we use a 3-level scheme (discuss with Jonathan).

The time evolution of the system is based on the assumption that the probability of an event occuring is independent of the history of the system and therefore it obeys Poisson statistics. This assumption will be justified below in Subsection \ref{subsec:justification}. Following Ref.\cite{ref:bortz1975}, the probability that an event $i$ occurs in the infinitessimal time interval $(t,t+dt)$ is $p_i(t)dt = r_idt$ where $r_i$ is the rate associated with the event. The total probability $P(t)$ that an event occurs in time interval  $(t,t+dt)$ is then

\begin{equation}
P(t)dt = \sum_i \, p_i(t)dt = \sum_i \, r_idt = Rdt.
\end{equation}
Let $\pi(t)$ be the probability that no event occurs in time interval $(0,t)$. Then, the probability that no event occurs in time interval $(0,t+dt)$ is equal to the probability that no event has occured in time interval $(0,t)$ and the probability than no event occurs in time interval $(t,t+dt)$, namely,

\begin{equation}
\pi(t+dt) = \pi(t)(1 - Rdt) = \pi(t) -\pi(t)Rdt.
\end{equation}
After rearranging and taking the limit $dt\rightarrow0$, we obtain

\begin{equation}
\frac{d\pi(t)}{dt} = \displaystyle\lim_{dt\to0}\frac{\pi(t+dt)-\pi(t)}{dt} = -\pi(t)R
\end{equation}
which can be easily solved to give

\begin{equation}
\pi(t) = \pi(0)e^{-Rt} \label{eq:time}
\end{equation}
with $\pi(0) = 1$ as the probability of no event occuring in time $t = $ 0 is unity. The physical time of each time step in the simulation is then assumed to be distributed as Eq.(\ref{eq:time}) and a second random number $\rho_2 \in (0,1)$ is chosen such that the physical time $\tau$ is given via $\rho_2 = e^{-R\tau}$ or

\begin{equation}
\tau = -\frac{\ln(\rho_2)}{R}.
\end{equation}
This expression gives an average time between events equal to $1/R$.

\subsection{Use of kinetic Monte Carlo models to describe epitaxial growth} \label{subsec:justification}

In this section we motivate the use of a stochastic Monte Carlo technique for the description of epitaxial growth systems.

As described in Ref.\cite{ref:evans2006}, deposition often occurs by the random impingement of atoms or molecules on a substrate. Within a KMC simulation, a deposition flux is included in the model, and the deposition sites are chosen randomly, in accordance with experimental realisations.

After deposition, atoms are adsorbed on specific sites on the lattice that correspond to minima in energy for the configuration of the system. A given configuration of the adatoms on the substrate is called a \textit{state of the system}. The adatoms vibrate at frequencies of the order $\nu_0\sim10^{13}$ s$^{-1}$ about the minima they occupy. However, these vibrations do not change the state of the system, only the state of the individual adatoms. The state of the system changes occasionally when an adatom vibration makes the adatom move from one minimum to another by going past an energy barrier $E$. Such transitions are thermally activated, so the rate $r$ associated with them is of the Arrhenius form.

A KMC model to describe epitaxial growth takes advantage of the different time scales between the change of state of an adatom and of the entire system by only considering the latter. During the irrelevant vibrations of an adatom about a given minimum, the system loses memory about the previous configurations it occupied, and therefore one can use Poisson statistics to describe a KMC time step.

\chapter{Standard kinetic Monte Carlo model of epitaxial growth} \label{ch:standard}

In this chapter we introduce what we call the KMC \textit{standard model} of epitaxial growth. The objective is two-fold: first, to provide an example of the N-fold way algorithm described above; and second, to have a simple model that is going to serve as a starting point for further studies and as a reference to compare with the graphene system. 

\section{Description}
We consider a 200$\times$200 square lattice with periodic boundary conditions. The model describes two processes: deposition of single atoms and diffusion of adatoms (adsorbed atoms) over the substrate. This model does not aim to describe the graphene system, but the studies we will perform will be keeping in mind the final objective to study epitaxial graphene growth. 

\textit{Deposition} occurs at a constant flux $F$, and atoms can only be deposited at an unoccupied lattice location. In the case that a deposition attempt is made on an occupied site, then a deposition process is still forced to happen in order to keep the deposition flux constant, and a different site is chosen repeatedly until an unoccupied site is found. The total coverages explored here are a small fraction of the lattice because we are interested in the initial stages of epitaxial graphene growth. Then, such events are rare indicating that the above modelling of deposition does not introduce an undesired effect in the model.  

\textit{Diffusion} occurs between nearest-neighbour sites, and an adatom can only diffuse to an unoccupied site. The diffusion rate $D$ is given by the Arrhenius form

\begin{equation}
D = \nu_0e^{-(E_a+nE_b)/k_BT}
\end{equation}
where $\nu_0 = 2k_BT/h$ for Boltzmann constant $k_B$, temperature $T$ and Planck constant $h$. Furthermore, $E_a$ is the energy barrier associated with the diffusion of an adatom, and $E_b$ is an extra energy barrier associated with a bond to a nearest neighbour. The variable $n$ is the number of nearest neighbours of the diffusing adatom, and the above \textit{bond-counting} model is typically used in epitaxial growth simulations \cite{ref:evans2006}. Note this rate depends on the local configuration around the diffusing adatom, where nearest-neighbours result in a penalty energy for diffusion. That is, adatoms that share bonds with others are less likely to diffuse away and this is the key ingredient to get island formation. 

Temperature is the parameter we tune in order to explore the different regimes of the system. The energy barrier for adatom diffusion is kept fixed at $E_a =$ 1.00 eV, while the bond energy barrier $E_b$ is varied. The flux is set to $F =$ 1 ML/s and the total coverage reached is $\theta =$ 0.1 ML. The units of coverage are monolayers (ML) where one monolayer corresponds to a layer of epitaxial material grown on the substrate.

The model presented here belongs to a class of models referred to as generic models \cite{ref:evans2006}. These generic models do not have any specifications of critical island size, that is, no island size above which attachment is irreversible is defined. Instead, a set of local rules such as the bond-counting scheme above is defined and the system is left to evolve according to these only.

\section{Results and discussion}
In this section we analyse the physical processes involved in epitaxial growth based on the KMC model described above. We start with a qualitative look at the growth behaviour, and then we move on to a quantitative analysis that will serve as a basis for the later study of the graphene system.

\subsection{Physical lattice}

\begin{figure}[h!]
\centering
\includegraphics[scale=1.6]{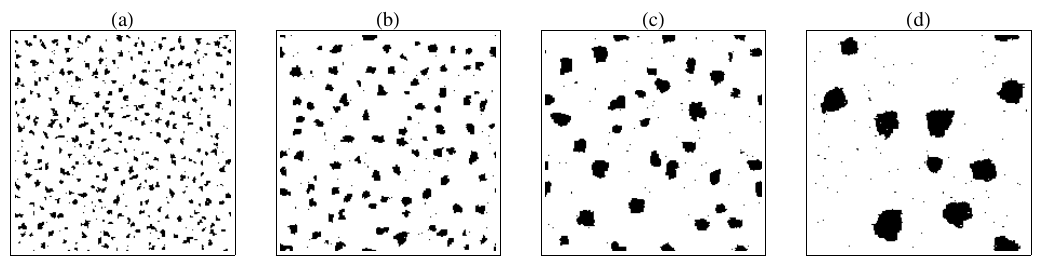}
\caption{\label{fig:standardlattice}Physical lattice at temperatures (a) $T =$ 550 K, (b) $T =$ 600 K, (c) $T =$ 650 K, (d) $T =$ 700 K. In all cases, $E_a = 1.00$ eV, $E_b = 0.20$ eV, $F = 1$ ML/s and $\theta = 0.1$ ML. Black squares represent occupied sites and white squares unoccupied sites.}
\end{figure}

In Fig. \ref{fig:standardlattice} we can see the physical lattice for different temperatures. All instances are after the last adatom of a total coverage of $\theta =$ 0.1 ML has been deposited, corresponding to a 10\% coverage of the lattice. The bond energy barrier used is $E_b =$ 0.20 eV.

It can be seen from Fig. \ref{fig:standardlattice} that as temperature increases the islands that are formed become larger in size and fewer in number. Adatoms at the edges of islands are more weakly bound to the island because they have a smaller number of bonds, so minimising their number is thermodynamically favourable. This leads to large and circular islands being the most stable configuration for the system because it minimises the interface free energy.

In Fig. \ref{fig:standardlattice}, larger temperatures present this thermodynamically favourable state. This is caused by the high thermal energies that lead to large adatom mobility and the system can easily reach the most favourable configuration. Under these conditions, islands that are not very large can break up and only a few nucleations will eventually lead to stable islands, which will be fewer in number. Also, the islands will be larger in size because there will be less competition for adatom incorporation due to the presence of a smaller number of competing islands.  

When the flux is turned off all the different cases depicted in Fig. \ref{fig:standardlattice} will tend to configurations minimising the free energy for large times, even when the thermal energies do not facilitate it. A common mechanism that leads to coarsening in submonolayer epitaxial systems is Ostwald ripening \cite{ref:lifshitz,ref:petersen2003}.

\begin{figure}[h!]
\centering
\subfloat[Adatom density]{\includegraphics[width=0.45\textwidth]{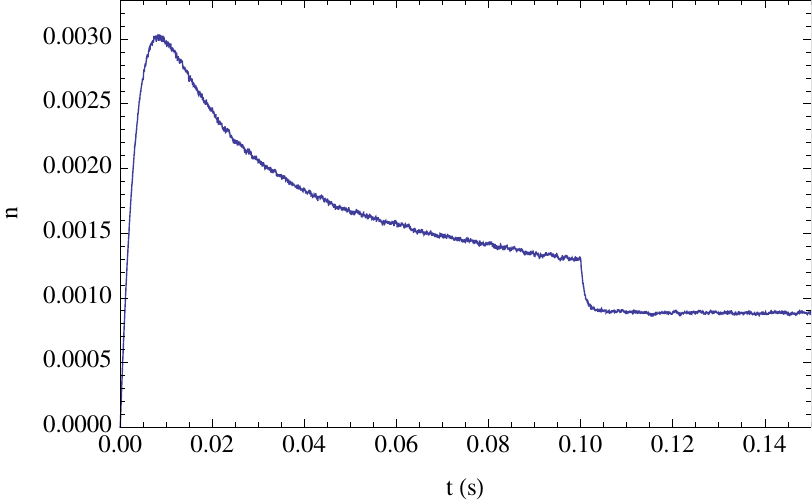}}
\hspace{0.04cm}
\subfloat[Island density]{\includegraphics[width=0.45\textwidth]{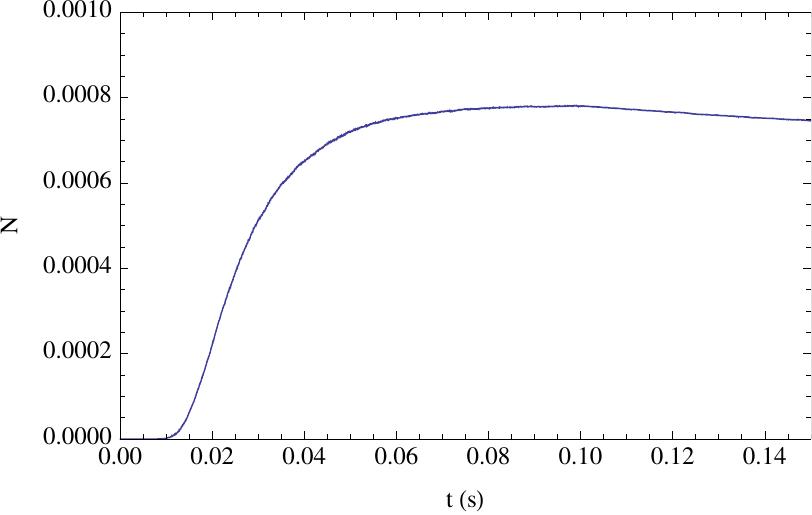}}\\
\caption{\label{fig:tdependence}Adatom density $n$ and island density $N$ as a function of time. The system corresponds to the standard KMC model with bond energy $E_b =$ 0.20 eV and $T =$ 650 K.}
\end{figure}

These features can be studied quantitatively by looking at the island size distribution and at the time and temperature dependence of adatom and island densities as described in the following subsections. For clarity, we plot in Fig. \ref{fig:tdependence} the time dependence of both adatom and island densities for the parameters $E_b =$ 0.20 eV and $T =$ 650 K. Such profiles are used in the various analyses presented below for the island size distribution, the adatom density at nucleation $n_{nuc}$, the adatom density at equilibrium $n_{eq}$ and the island density at equilibrium $N_{eq}$. In the instances shown in Fig. \ref{fig:tdependence}, the deposition flux is turned off at a coverage of $\theta =$ 0.10 ML (corresponding to $t =$ 0.10 s for $F =$ 1 ML/s) and the system is then allowed to relax for a further 0.05 seconds. The main features discussed in the previous chapter for such profiles are clearly observed.

\subsection{Island size distribution}

The island size distribution is a quantity readily available in KMC studies, but difficult to obtain in mean field approaches such as rate equations.

\begin{figure}[h!]
\centering
\includegraphics[scale=1.4]{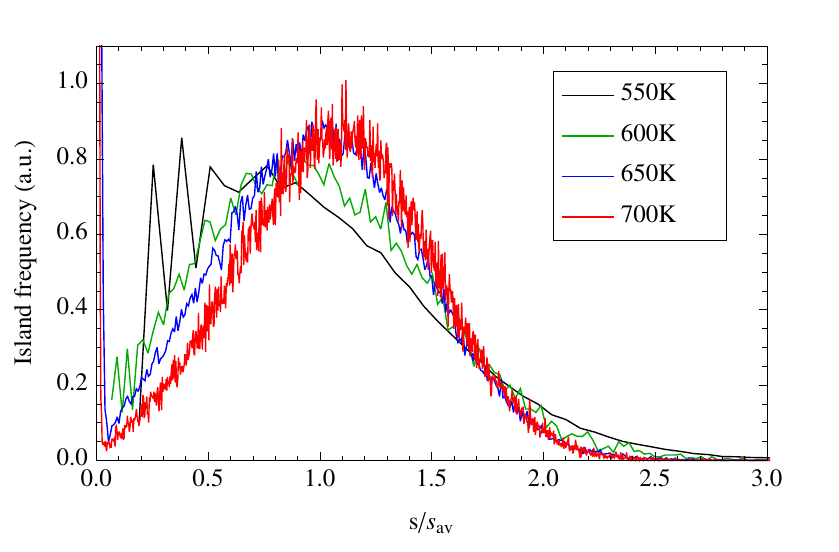}
\caption{\label{fig:distributions}Island size distribution for the standard KMC model with bond energy $E =$ 0.20 eV.}
\end{figure}

In Fig. \ref{fig:distributions} we plot the island size distribution as it is at a coverage of $\theta$ = 0.1 ML, for bond energy $E_b =$ 0.20 eV and for a range of temperatures. The distributions are calculated when the system has been allowed to relax after turning the deposition flux off. The island sizes are normalised in such a manner that the area under the curve is unity, and the horizontal axis represents $s/s_{av}$ for island size $s$ and average island size 

\begin{equation}
s_{av} = \frac{\sum_s sN_s}{\sum_s N_s},
\end{equation}
where $N_s$ is the number of islands of size $s$. 

It can be seen in Fig. \ref{fig:distributions} that as temperature increases the distribution becomes narrower. At large temperatures, islands can break up more easily than at low temperatures, so we expect that there is some correction in island sizes and all islands approach an ideal size. This analysis quantifies the discussion above based on snapshots of the lattice in Fig. \ref{fig:standardlattice}.

To exemplify this island size correction at high temperatures consider a system where two islands form nearby. At low temperatures, the islands will grow by attachment of adatoms, but because they are nearby they compete for free adatoms in the surface, so both islands will in general be smaller than the average island size. However, the same situation at high temperatures can result in the break-up of one of the two islands, and the incorporation of its adatoms into the other, so that only one island is left, and its size is closer an average size. Note that in the same manner, if an island is formed in a region where it is isolated, for low temperatures it will incorporate all adatoms in its vicinity and thus grow to larger than average sizes. However, at large temperatures the detachment of adatoms from the island will be more significant and the size of the island will again be closer to that of others.

The data for island size distribution is not as clear as the data for adatom or island concentrations, specially for large temperatures. This is because for high temperatures very few islands are formed on the lattice, in Fig. \ref{fig:standardlattice} only 10 islands for $T =$ 700K. Furthermore, the islands formed are larger in size, so the range of possible island sizes is broader. These two characteristics mean that to obtain useful data for island size distributions we need to calculate an average over many different instances of the system evolution. For instance, in the case depicted in Fig. \ref{fig:distributions} for $T= $ 700K, up to 10,000 instances were used.

\subsection{Adatom density}

The experimental results reported by Loginova and co-workers \cite{ref:loginova2008,ref:loginova2009} represent some of the cleanest data on adatom density as a function of time for epitaxial growth experiments. As described above, the adatom density presents interesting features, and we will concentrate on its values at the onset of nucleation, $n_{nuc}$, and at equilibrium with islands, $n_{eq}$.

\begin{figure}
\centering
\includegraphics[scale=1.6]{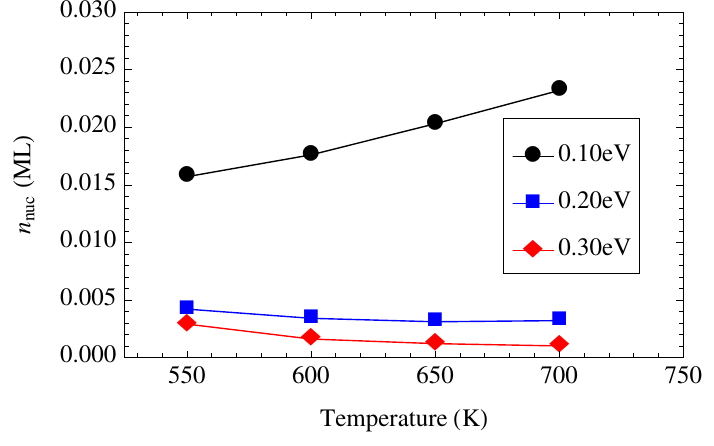}
\caption{\label{fig:standard}Adatom nucleation density $n_{nuc}$ as a function of temperature for nearest neighbour bond strengths $E_b = $ 0.10, 0.20, 0.30 eV.}
\end{figure}

The behaviour of $n_{nuc}$ is highly dependent on the strength of the bond between adatoms as shown in Fig. \ref{fig:standard}. For low bond energy the critical island size is large, that is, the size required for a cluster of atoms to eventually become an island is large because small clusters will break-up with large probability due to the weakness of the bonds keeping the atoms together. Furthermore, as temperature increases bonds break more often, so we expect that with increasing temperature nucleation occurs at larger adatom concentrations. In contrast, for large bond energies the thermal vibrations cannot easily break the bonds, and the critical island size is reduced to a few atoms even for high temperatures. Then the limiting process for nucleation is adatom diffusion because as soon as adatoms become nearest neighbours nucleations occur. Therefore, as temperature is increased, the larger diffusion of adatoms means that the adatom nucleation occurs sooner and $n_{nuc}$ decreases with temperature. 

In Fig. \ref{fig:standard} we plot $n_{nuc}$ as a function of temperature, and for different adatom bond energies $E_b$. Each data point has a statistical error associated with it that is of order $\sim$0.01 of the value plotted so it is not depicted. We can observe both the low bond energy limit with $E_b =$ 0.10 eV and the large bond energy limit with $E_b = $ 0.30 eV. With bond energy $E_b =$ 0.20 eV we can see an intermediate scenario where the lower temperature range is diffusion dominated and the adatom nucleation concentration decreases with temperature, but at larger temperatures island break-up becomes dominant and a larger adatom concentration is required for nucleation because the critical island size increases. Note also that as expected, for lower bond energies $n_{nuc}$ is much larger than for large bond energies.

\begin{figure}[h!]
\centering
\includegraphics[scale=1.6]{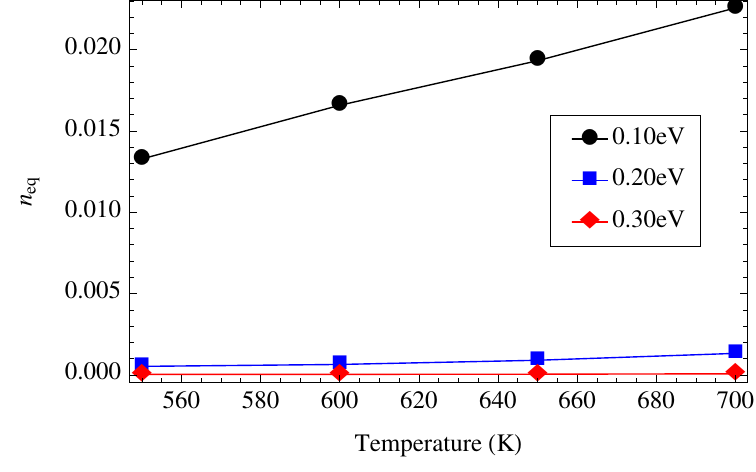}
\caption{\label{fig:standardeq}Adatom equilibrium density $n_{eq}$ as a function of temperature for nearest neighbour bond strengths $E_b = $ 0.10, 0.20, 0.30 eV.}
\end{figure}

The adatom density at equilibrium is shown in Fig. \ref{fig:standardeq}, plotted as a function of temperature and for different $E_b$. In all cases the equilibrium concentration increases with temperature because detachment from islands dominates at larger temperatures. For low binding energy $E_b =$ 0.10 eV, $n_{eq}$ is large because detachment rates are very large. In contrast, for higher bond energies the adatoms do not detach from islands with such large rates and $n_{eq}$ almost vanishes. For the lower temperatures and higher binding energies equilibrium is only reached after a long time because low diffusion over the substrate means adatoms do not attach to islands frequently, but strong bonds lead to large islands, so the system takes a long time to equilibrate.

\subsection{Island density}

In this subsection we study the equilibrium island density $N_{eq}$ for the standard model. Experimental data on island density is available for the graphene system, and the RE approach fails to describe it. Therefore, a careful analysis of island density within KMC could deepen our understanding of the growth kinetics of graphene.

In the bond-counting scheme used for the standard model, a critical island size is not well-defined. Therefore, the definition of island is not very clear. For instance, for low bond energies between nearest neighbours, dimers are very unstable and break-up frequently. However, for large bond energies, dimers are very stable and lead to large islands in most cases. In the standard model, clusters of size larger than 10 are treated as islands. This choice might not be strictly accurate, but the trends to be observed in the data arise clearly, therefore simplicity has prevailed in the choice.

\begin{figure}[h!]
\centering
\includegraphics[scale=1.6]{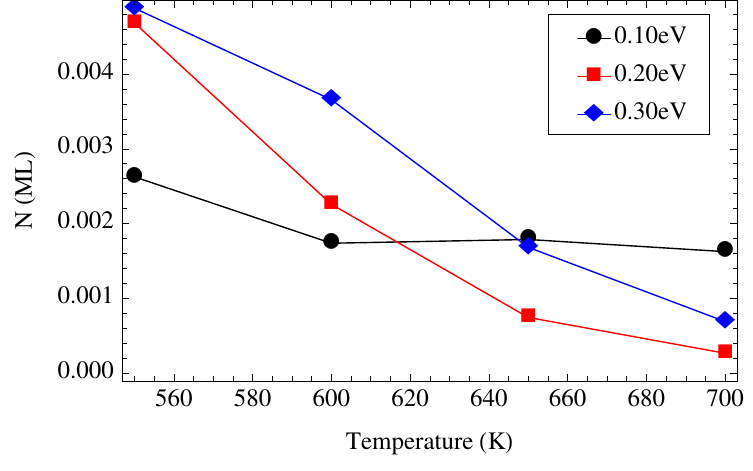}
\caption{\label{fig:standardidensity}Equilibrium island density $N_{eq}$ as a function of temperature for nearest neighbour bond strengths $E_b = $ 0.10, 0.20, 0.30 eV.}
\end{figure}

\begin{figure}[h!]
\centering
\includegraphics[scale=1.6]{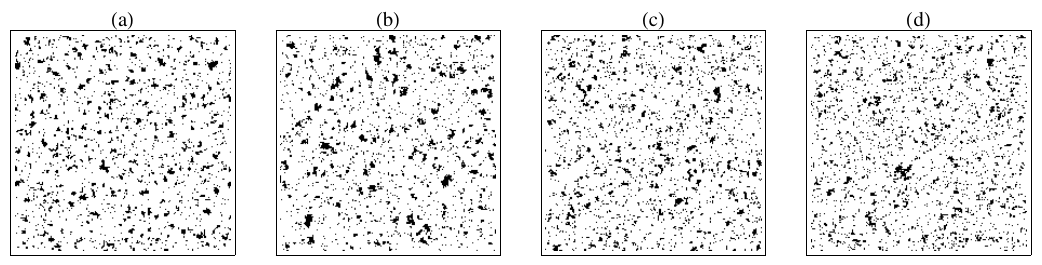}
\caption{\label{fig:standardlattice10}Physical lattice at temperatures (a) $T =$ 550 K, (b) $T =$ 600 K, (c) $T =$ 650 K, (d) $T =$ 700 K for a bond energy of $E_b =$ 0.10 eV. Black squares represent occupied sites and white squares unoccupied sites.}
\end{figure}

In Fig. \ref{fig:standardidensity} we plot the equilibrium island density as a function of temperature. For the stronger bond cases $E_b =$ 0.20, 0.30 eV, a clear decreasing trend for $N_{eq}$ with temperature is observed. This is in agreement with the physical lattices above. Note also that the concentration of islands is larger for the case with stronger binding. This is caused by the higher difficulty of small clusters to break up if bonds are strong, and therefore more nucleations are expected in this regime. The low bond energy case $E_b =$ 0.10 eV presents a different behaviour. For the temperature increase from 550 K to 600 K, a decrease in $N_{eq}$ is observed as expected. However, further temperature increases do not lead to a clear decrease, instead the island concentration remains approximately constant. This behaviour is very particular of the low bond energy case where islands cannot grow very large due to the high adatom detachment rates. The observed behaviour of island density means that already at $T\sim$ 600 K the adatom density on the substrate determines the minimum island density attainable with the high adatom detachment rate present. Then, further temperature increases do not lead to a smaller island concentration. This behaviour is confirmed by the physical lattice in this regime shown in Fig. \ref{fig:standardlattice10}, where temperatures higher than $T =$ 600 K do not change the aspect of the system as significantly as for the case depicted above in Fig. \ref{fig:standardlattice}. The islands decrease slightly in size due to the higher temperatures causing more frequent bond breaking.

The behaviour of the system suggests that we use a nearest neighbour bond energy of the order $E_b\sim$ 0.20 eV or above for further studies of graphene. This is because lower binding energies will not lead to growth of large islands. 

\subsection{Error analysis}

The different values presented above have a statistical uncertainty associated with them due to the stochastic nature of the KMC method. These statistical errors are minimised by averaging the results over a large number of experiments, and this has been done in a manner that the errors of a given quantity are about 1\% of the value of the quantity throughout this work. These small statistical errors are not shown in the plots because they are too small to be seen appropriately. In the quantity $N_{eq}$ the statistical errors can be up to 10\% of the value quoted, but even in these circumstances the trends discussed can be clearly observed.

Another error consideration comes in when evaluating the correctness of the model and its implementation. The model described here is a simple bond-counting epitaxial growth model that is widely used in such studies. Its implementation has been partially checked by finding the expected results for such bond-counting model. In the next chapter we consider this point further for the KMC model to describe epitaxial graphene growth.

\chapter{Epitaxial graphene growth kinetic Monte Carlo model} \label{ch:tetramer}

In this chapter we present a KMC model for the study of the kinetics of epitaxial graphene growth. The experimental observations reported about the system \cite{ref:loginova2008,ref:loginova2009} are explained with the model, and the relevant physical processes determining the observed behaviour are identified.

However, direct comparison with experiment is not possible for two reasons. First, the values of the energy barriers associated with the different processes are not available yet. The present work determines the relevant physical processes to include in the study of epitaxial graphene growth. Therefore, the next step is to determine the exact energetics of these identified processes, for instance using \textit{ab initio} techniques. Second, the deposition flux used experimentally is very low, resulting in time scales that are computationally inaccessible to the present KMC model.

\section{Description}

%An extension to the dimer model presented in Section SOMETHING is described in this chapter. The objective is to have a KMC model which closely resembles the known graphene growth system. With this extended model, we study the physical processes governing the different features found experimentally about the kintetics of epitaxial graphene growth.

We introduce a KMC model for epitaxial graphene growth in a 200$\times$200 square lattice with periodic boundary conditions. We include adatoms, islands and an intermediate species formed by four adatoms arranged in a square which we call \textit{tetramers}. Experiments indicate that the intermediate species in the graphene system are 5-atom clusters, but for convenience with the underlaying square lattice we choose the simpler symmetric tetramer. The presence of 5-clusters in graphene growth is most probably determined by the stability of such structures on the substrates (observed in grand canonical Monte Carlo simulations for Niquel Ni(111) \cite{PhysRevB.73.113404}), but we postulate that the kinetics are determined only by the presence of this intermediate species rather than by its details. This justifies the use of tetramers which represent the stable symmetric structure in our model. The validity of this assumption can be confirmed later with the results obtained. In a similar manner, the use of a square lattice rather than a triangular lattice can be justified for simplicity of the model. 

Deposition occurs as described above for the standard model via individual atoms. Experimentally, both carbon atoms and ethylene molecules are deposited, but the behaviour of the growth system is independent of the deposited species \cite{ref:loginova2009} and for simplicity we only consider individual atom deposition.

Adatoms diffuse over the substrate in the same manner as for the standard model. They only interact to form tetramers and to attach to islands. A tetramer is formed whenever four adatoms come together in a square configuration on the lattice. From that moment onwards, the tetramer is treated as an individual species formed by four occupied sites in a square configuration. Attachment of adatoms to islands occurs as described above for the standard model whenever an adatom diffuses to a site with nearest neighbours belonging to an island. After an adatom has attached to an island, it is still treated as an individual atom, but subject to the bond-counting scheme described above. We stress this only occurs for adatoms that diffuse to sites with nearest neighbours belonging to islands: if a tetramer is a nearest neighbour, there is no interaction with the adatom based on the fact that tetramers are a very stable and independent species.

\vspace{3cm}

\begin{figure}[h!]

\begin{picture}(0,0)

%Figure 1
\put(0,40){(a)}

\put(20,0){\line(1,0){60}}
\put(20,20){\line(1,0){60}}
\put(20,40){\line(1,0){60}}
\put(20,60){\line(1,0){60}}
\put(20,80){\line(1,0){60}}
\put(20,0){\line(0,1){80}}
\put(40,0){\line(0,1){80}}
\put(60,0){\line(0,1){80}}
\put(80,0){\line(0,1){80}}

\put(50,30){\circle*{10}}
\put(50,50){\circle*{10}}

%Figure 2
\put(110,40){(b)}

\put(130,0){\line(1,0){80}}
\put(130,20){\line(1,0){80}}
\put(130,40){\line(1,0){80}}
\put(130,60){\line(1,0){80}}
\put(130,80){\line(1,0){80}}
\put(130,0){\line(0,1){80}}
\put(150,0){\line(0,1){80}}
\put(170,0){\line(0,1){80}}
\put(190,0){\line(0,1){80}}
\put(210,0){\line(0,1){80}}

\put(160,50){\circle*{10}}
\put(180,50){\circle*{10}}
\put(160,30){\circle*{10}}
\put(180,30){\circle*{10}}

%Figure 3
\put(240,40){(c)}

\put(260,0){\line(1,0){100}}
\put(260,20){\line(1,0){100}}
\put(260,40){\line(1,0){100}}
\put(260,60){\line(1,0){100}}
\put(260,80){\line(1,0){100}}
\put(260,0){\line(0,1){80}}
\put(280,0){\line(0,1){80}}
\put(300,0){\line(0,1){80}}
\put(320,0){\line(0,1){80}}
\put(340,0){\line(0,1){80}}
\put(360,0){\line(0,1){80}}

\put(290,30){\circle*{10}}
\put(290,50){\circle*{10}}
\put(310,30){\circle*{10}}
\put(310,50){\circle*{10}}
\put(330,30){\circle*{10}}
\put(330,50){\circle*{10}}

\put(390,40){\circle*{10}}
\put(405,37){tetramer}

\end{picture}

\vspace{0.3cm}

\caption{\label{fig:nucleation}Schematic of possible nucleation configurations for (a) $j =$ 2, (b) $j =$ 4 and (c) $j =$ 6, where $j$ is the number of tetramers. For the asymmetric cases $j = $ 2, 6; a $\pi/2$ rotation with respect to the cases depicted is also a valid nucleation configuration. Note that in the diagram a black circle represents four adatoms in a square configuration and a square lattice site represents four square lattice sites in the actual simulation.}

\end{figure}
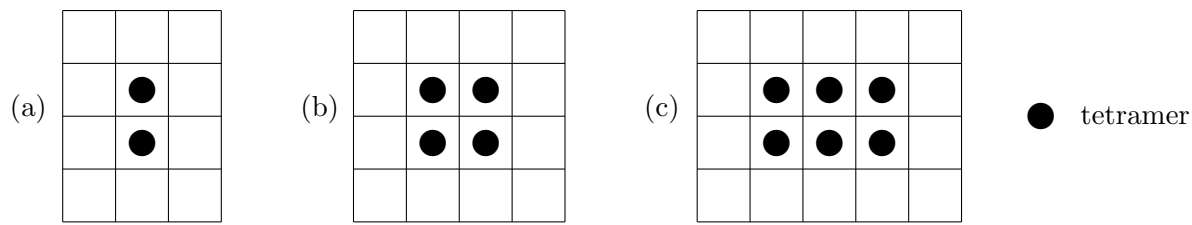

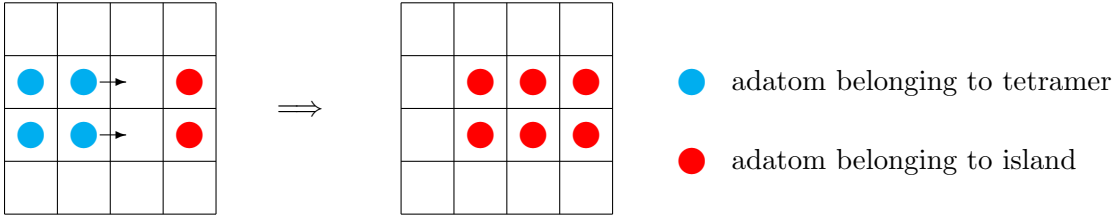
\begin{figure}[h!]

\vspace{3cm}

\begin{picture}(0,0)

%Figure 1
\put(20,0){\line(1,0){80}}
\put(20,20){\line(1,0){80}}
\put(20,40){\line(1,0){80}}
\put(20,60){\line(1,0){80}}
\put(20,80){\line(1,0){80}}
\put(20,0){\line(0,1){80}}
\put(40,0){\line(0,1){80}}
\put(60,0){\line(0,1){80}}
\put(80,0){\line(0,1){80}}
\put(100,0){\line(0,1){80}}

\color{cyan}
\put(50,30){\circle*{10}}
\put(30,30){\circle*{10}}
\put(30,50){\circle*{10}}
\put(50,50){\circle*{10}}
\color{black}

\color{red}
\put(90,30){\circle*{10}}
\put(90,50){\circle*{10}}
\color{black}

\put(56,30){\vector(1,0){10}}
\put(56,50){\vector(1,0){10}}

%\put(120,40){\vector(1,0){30}}
\put(120,37){$\implies$}

%Figure 2
\put(170,0){\line(1,0){80}}
\put(170,20){\line(1,0){80}}
\put(170,40){\line(1,0){80}}
\put(170,60){\line(1,0){80}}
\put(170,80){\line(1,0){80}}
\put(170,0){\line(0,1){80}}
\put(190,0){\line(0,1){80}}
\put(210,0){\line(0,1){80}}
\put(230,0){\line(0,1){80}}
\put(250,0){\line(0,1){80}}

\color{red}
\put(200,50){\circle*{10}}
\put(220,50){\circle*{10}}
\put(200,30){\circle*{10}}
\put(220,30){\circle*{10}}
\put(240,30){\circle*{10}}
\put(240,50){\circle*{10}}
\color{black}

\color{cyan}
\put(280,50){\circle*{10}}
\color{black}
\put(295,47){adatom belonging to tetramer}

\color{red}
\put(280,20){\circle*{10}}
\color{black}
\put(295,17){adatom belonging to island}

\end{picture}

\vspace{0.3cm}

\caption{\label{fig:tetramer}Schematic of tetramer diffusion and attachment to islands.}

\end{figure}

Tetramers can diffuse over the substrate, in a manner analogous to adatom diffusion. The only difference is that the four adatoms forming the tetramer move at the same time and in the same direction as an individual unit. Tetramers only interact with other tetramers to nucleate and with islands to attach to them. Nucleation occurs when a number $j$ of tetramers come together on the lattice in a symmetric manner as indicated in Fig. \ref{fig:nucleation}, where three possible values for $j$ are shown. As soon as a nucleation occurs, the tetramers lose their identity and the adatoms forming them become members of an island and are subject to nearest neighbour interactions via the bond-counting scheme. Tetramers can also attach to existing islands, and such events occur whenever the two atoms of a side of the tetramer become nearest neighbours to sites occupied by islands, in such a manner that a side of the tetramer square is in contact with an island. This is schematised in Fig. \ref{fig:tetramer}. When a tetramer attaches to an island, it again loses its identity and the individual atoms become subject to nearest neighbour interactions.

%\vspace{3cm}

We tune the temperature whilst keeping the energy barriers shown in Table \ref{tab:gparameters} at fixed values. Note the tetramer diffusion barrier is kept lower than the adatom diffusion barrier to promote tetramers to the dominant species in graphene growth kinetics. 

\begin{table}[h!]
\centering
\begin{tabular}{lll}
\hline
\textbf{Description} & \textbf{Parameter} & \textbf{Value}\\ 
\hline
Coverage & $\theta$ & 0.10 ML\\
Adatom diffusion barrier & $E_a$ & 1.00 eV\\
Tetramer diffusion barrier & $E_t$ & 0.80 eV\\
Adatom bond energy & $E_b$ & 0.20 eV\\
\hline
\end{tabular}
\caption{Values of the basic parameters used in the KMC model to describe epitaxial graphene growth} \label{tab:gparameters}
\end{table}

The model presented here belongs to the generic models class. A possible variation of this model for the simulation of epitaxial graphene growth is described in Appendix \ref{app:growth}.

In the following section we investigate how different processes superimposed on top of this basic model determine the rich behaviour observed experimentally for the kinetics of epitaxial graphene growth.

\section{Results and discussion}

In this section we study the three experimental observations by Loginova and co-workers on the epitaxial graphene growth system \cite{ref:loginova2008,ref:loginova2009}. For convenience, we repeat them here:

\begin{enumerate}
\item The adatom density at the onset of nucleation increases with increasing temperature.
\item The adatom density at equibrium $n_{eq}$ is roughly half of the adatom density at the onset of nucleation $n_{nuc}$, namely, $n_{nuc} \sim 2n_{eq}$.
\item The island density at equilibrium decreases rapidly with increasing temperature.
\end{enumerate}
In the following subsections we study them individually. There is only preliminary data for the island size distribution, and it is presented in Appendix \ref{app:isd}.

\subsection{Adatom density at nucleation}

It was discussed in Section \ref{subsec:exp}  that the adatom density at nucleation, $n_{nuc}$, is experimentally found to increase as a function of temperature for the graphene growth system. In standard growth systems, we found in Chapter \ref{ch:standard} that $n_{nuc}$ could present different behaviours with temperature depending on the binding energy between adatoms. In this section we will investigate this effect on the KMC graphene model and find a more complex behaviour. 

Inspired by experiments \cite{ref:cui2011,PhysRevLett.103.166101} and theoretical studies \cite{ref:zangwill}, it appears that for graphene the critical nucleus sizes are of the order of tens of adatoms. In our KMC model we have incorporated this observation by allowing nucleations to occur only when a certain number $j$ of tetramers come together in specific configurations. Because islands only start appearing with these large sizes, then the analysis on the standard model with low binding energy might not be applicable for graphene. This is because even for low bond energies, the incipient islands are so large that dissociation might be very unlikely and the growth process would be diffusion limited. Such a scenario would lead to a decrease in $n_{nuc}$ with increasing temperature, even for low energy bonds. Furthermore, it was concluded in the previous chapter that binding energies which are too low do not result in the appearence of large islands. 

Even though we might have a diffusion limited growth system, epitaxial graphene growth is dominated by an intermediate stable species, and this difference with other growth systems can explain the increase of $n_{nuc}$ with temperature, even with the presence of relatively large incipient islands. Physically, a process is needed by which the concentration of clusters giving nucleation is decreased with increasing temperature to compensate for the larger cluster formation and diffusion rates. Such a process could be temperature dependent cluster break-up, included in the KMC model with a rate in the usual Arrhenius form subject to a cluster break-up energy barrier $E_k$.

\begin{figure}[h!]
\centering
\subfloat[Standard model]{\includegraphics[width=0.5\textwidth]{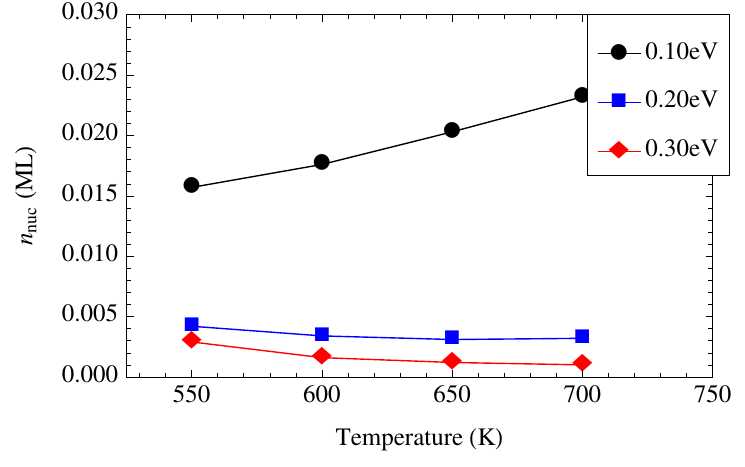}}
\subfloat[2-tetramer nucleation]{\includegraphics[width=0.5\textwidth]{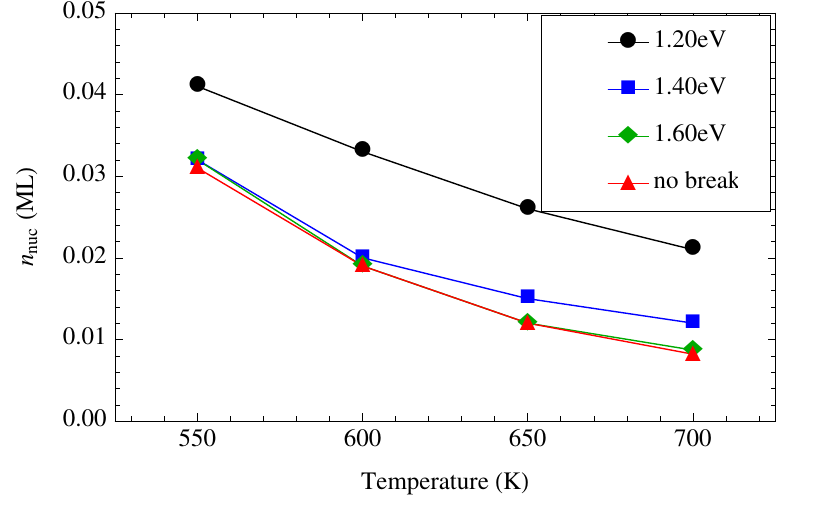}}\\
\subfloat[4-tetramer nucleation]{\includegraphics[width=0.5\textwidth]{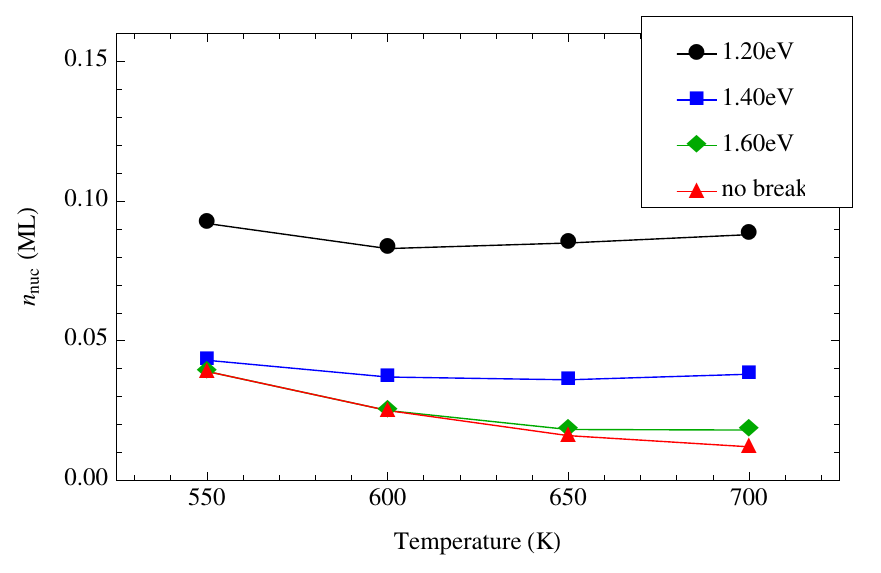}}
\subfloat[6-tetramer nucleation]{\includegraphics[width=0.5\textwidth]{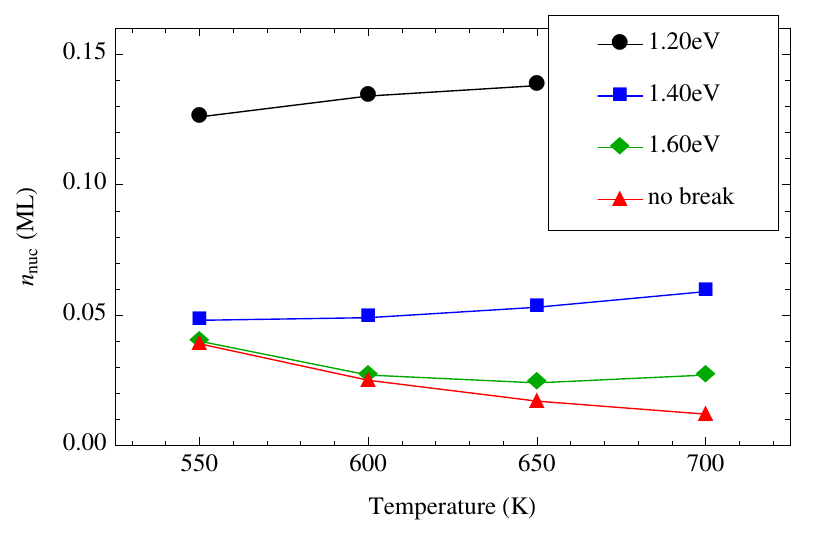}}
\caption{\label{fig:temp}Adatom density at the onset of nucleation as a function of temperature for (a) standard model, (b) nucleation with $j =$ 2, (c) nucleation with $j=$ 4, (d) nucleation with $j=$ 6. In the standard model different nearest neighbour bond energies are considered, and in the tetramer model different tetramer break-up energies are considered.}
\end{figure}

In Fig. \ref{fig:temp} we show the temperature dependence of $n_{nuc}$ for different nucleation sizes $j$ and for a range of cluster break-up energies $E_k$. We also plot the corresponding quantities for the standard model discussed above.

For $j = $ 2 we find that for all cluster break-up energy values, $n_{nuc}$ decreases with increasing temperature. This indicates that, as discussed above, an incipient island formed by $N =$ 8 adatoms is large enough not to dissociate, so that a low binding energy between adatoms cannot lead to an increase of $n_{nuc}$ with temperature as for the standard model even at high temperatures. Furthermore, cluster break-up is also found not to be sufficient to appropriately describe the temperature behaviour of $n_{nuc}$ of graphene growth. Even though the tetramer population is reduced as temperature increases, for $j = $ 2 nucleation is still a frequent event and the break-up rate is not high enough to act as a significant deterrent for nucleation. This means the system behaves as a typical diffusion limited growth system and $n_{nuc}$ decreases with temperature.

For $j =$ 4 more interesting behaviour starts appearing. For zero or very small cluster break-up, we still find a decreasing trend in $n_{nuc}$. Again, this is due to the behaviour as a typical diffusion limited growth system. However, as $E_k$ is decreased, an increase in $n_{nuc}$ is observed for the larger temperature ranges. With $j =$ 4, nucleation events are less likely than with $j =$ 2, so at large temperatures where cluster break-up is a significant effect, it can counteract the larger tetramer formation and diffusion rates. For a critical concentration of tetramers needed for nucleation, the high tetramer break-up rate means that larger adatom concentrations are needed to lead to the necessary tetramer concentration for nucleation. This leads to the increase of $n_{nuc}$ with temperature. It is not observed for low temperatures because diffusion still prevails over tetramer break-up in that regime.

For $j =$ 6, with the appropriate choice of $E_k$, an increase of $n_{nuc}$ with temperature is observed in the entire temperature range. This is the experimentally observed behaviour of epitaxial graphene growth. The large value of $j$ means that the tetramer concentrations needed for nucleation are so large that tetramer break-up can play a significant role even for low temperatures. From now on and unless otherwise stated we are going to use $j = 6$ and $E_k = 1.40$ eV for all further studies.   

\begin{figure}[h!]
\centering
%\subfloat[2-tetramer nucleation]{\includegraphics[width=0.3\textwidth]{./plots/tetramer2}}
\subfloat[4-tetramer nucleation]{\includegraphics[width=0.5\textwidth]{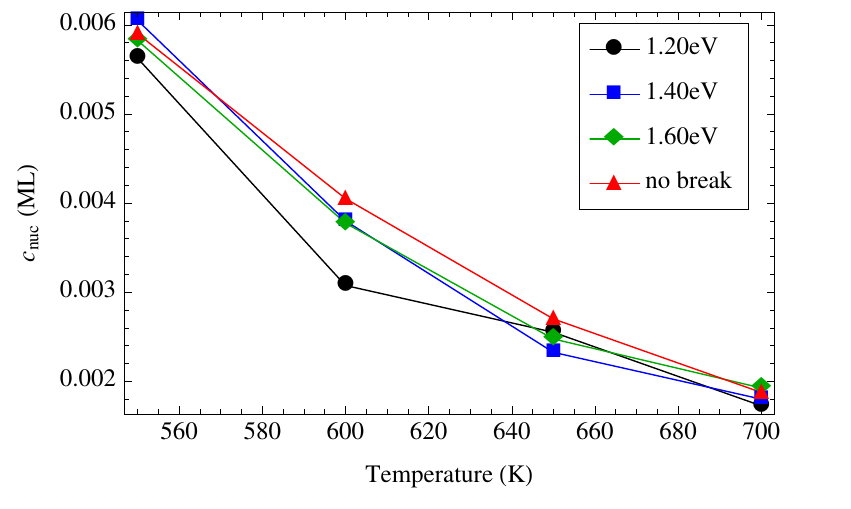}}
\subfloat[6-tetramer nucleation]{\includegraphics[width=0.5\textwidth]{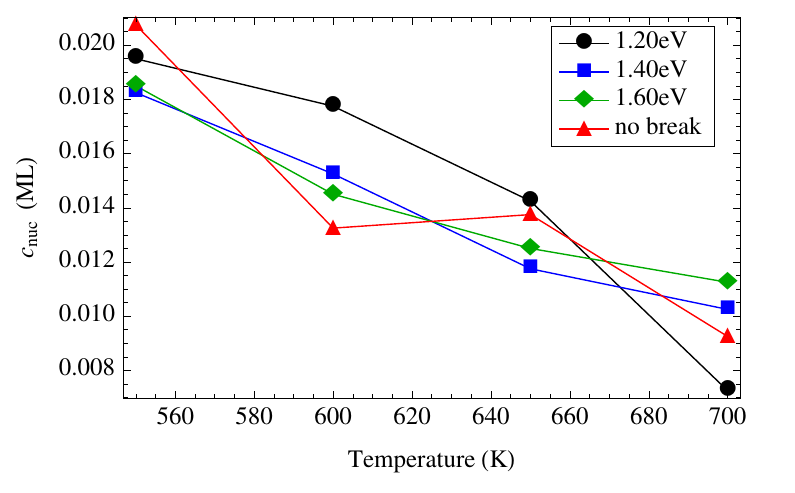}}
\caption{\label{fig:cnuc}Tetramer density at the onset of nucleation as a function of temperature for (a) nucleation with $j=$ 4 and (b) nucleation with $j=$ 6.}
\end{figure}

The above reasoning depends on the fact that the density of clusters at the onset of nucleation is independent of the cluster break-up energy barrier. This is indeed the case as shown in Fig. \ref{fig:cnuc} for $j = 4, 6$ and similar behaviour occurs for $j = 2$.

In agreement with the rate equations study by Zangwill and Vvedenksy \cite{ref:zangwill}, both cluster break-up and large incipient island sizes are found to be necessary to explain the increase of $n_{nuc}$ with temperature.

\subsection{Adatom density at equilibrium}

Adatom density at equilibrium is experimentally found to be roughly half of the adatom density at the onset of nucleation, $n_{nuc}\sim 2n_{eq}$. Such large $n_{eq}$ require the presence of a large attachment barrier for adatoms to graphene. The presence of such a barrier initially prompted the proposal of a 5-cluster attachment mechanism by Loginova and co-workers \cite{ref:loginova2008}. In the KMC model presented in this thesis, two physical processes determine the predominant growth of graphene by cluster attachment. The first one is a larger diffusion for tetramers than for adatoms, and the second is the presence of an adatom attachment barrier. It is only the second that determines the adatom concentration at equilibrium, and this effect is discussed here.

We consider a temperature dependent attachment barrier for adatoms to graphene. Computationally, it is incorporated in the KMC model via a Boltzmann factor $e^{-E_p/k_BT}$ such that when an adatom moves to a site with nearest neighbours belonging to an island, it will attach to the island with probability $p<e^{-E_p/k_BT}$. This condition makes adatom attachment more difficult at low temperatures than at high temperatures. Tuning the energy barrier  $E_p$ results in different values for $n_{eq}$. 

\begin{figure}[h!]
\centering
\includegraphics[scale=1.5]{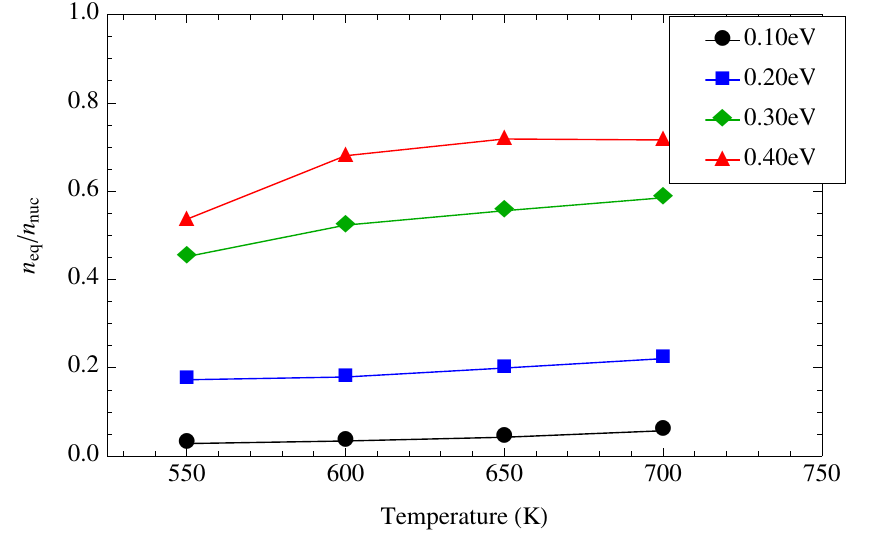}
\caption{\label{fig:adeq}Ratio of equilibrium adatom concentration $n_{eq}$ over adatom concentration at the onset of nucleation $n_{nuc}$ as a function of temperature. Different adatom attachment barriers are considered.}
\end{figure}

In Fig. \ref{fig:adeq} we plot the ratio $n_{eq}/n_{nuc}$ as a function of temperature for a range of adatom attachment barriers. We calculate $n_{eq}$ from the KMC model by turning the deposition flux off and allowing the system to relax, that is, by allowing the system to reach equilibrium between adatoms, tetramers and islands. As can be seen in Fig. \ref{fig:adeq}, the value of $E_p$ tunes the adatom equilibrium concentration, and typical experimental values $n_{nuc}\sim 2n_{eq}$ can be reached in our model for $E_p = $ 0.30 eV. The ratio is also seen to be roughly constant for the temperature range studied, although there is a slight increase with temperature. From now on unless otherwise stated we are going to take $E_p = 0.30$ eV.

Recalling $n_{eq}$ for the standard model in the previous chapter, we find significant differences. For the case with adatom bond energy of $E_b =$ 0.20 eV as used here, the standard model resulted in vanishingly small equilibrium concentrations. Therefore, it is clear that an adatom attachment barrier is essential in obtaining the correct behaviour.

The RE approach by Zangwill and Vvedensky \cite{ref:zangwill} also describes $n_{eq}$ accurately. However, in their model there is no adatom attachment barrier, so the only process that can also explain this behaviour is a large adatom detachment rate from islands. The lack of spatial information in the RE means that the capture zone of islands is smaller than in the KMC model, so adatoms attach to islands less frequently. This leads to higher values for $n_{eq}$ in the RE than in KMC models without an adatom attachment barrier.

\subsection{Island density at equilibrium}\label{subsec:Neq}

Island density at equilibrium is experimentally found to decrease rapidly with increasing temperature. In the KMC model used above for the description of graphene growth, as soon as a nucleation occurs, tetramers are more likely to attach to the existing island than to give rise to further nucleations. Therefore, to have a significant number of nucleations, the density of tetramers needs to be kept high by some mechanism. In order to try to describe this, we incorporate a tetramer attachment barrier $E_q$ in the same manner as an adatom attachment barrier above.

\begin{figure}[h!]
\centering
\includegraphics[scale=1.4]{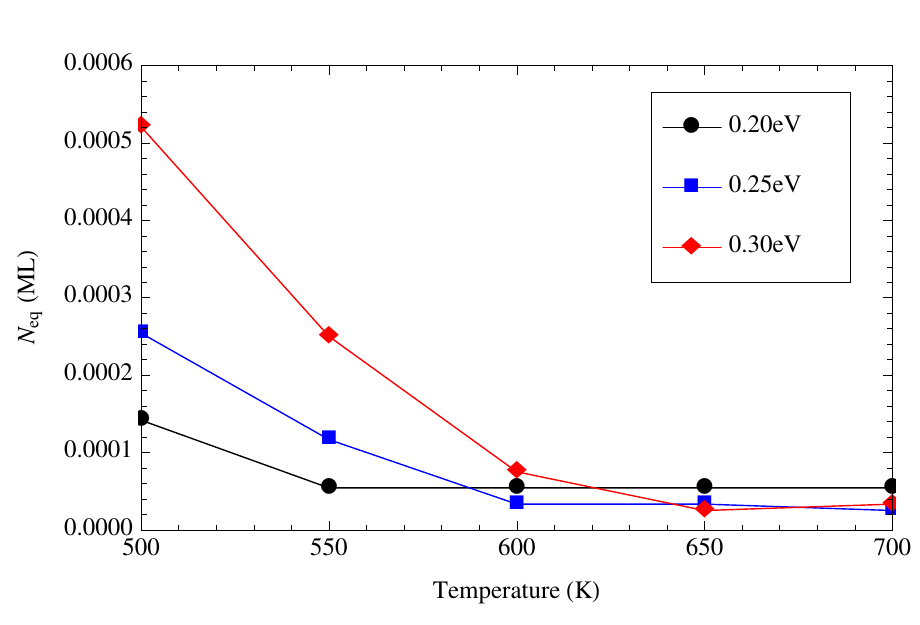}
\caption{\label{fig:iseq}Equilibrium island concentration $N_{eq}$ as a function of temperature. Different tetramer attachment barriers are considered.}
\end{figure}

In Fig. \ref{fig:iseq} we plot $N_{eq}$ as a function of temperature for a range of tetramer attachment barriers $E_q$. It can be seen that, by tuning the tetramer attachment barrier, a large decrease in island density can be engineered into the model. For the standard model discussed above, $N_{eq}$ decreased slowly as a function of temperature. For instance, for a bond energy $E_b =$ 0.20 eV, an increase of 100 K gave a decrease by a factor of 4. In the present case, an increase of 100 K for $E_q = 0.30$ eV gives an order of magnitude decrease in $N_{eq}$.  

Note that in order to get a large decrease in island density at nucleation, tetramer attachment barriers of the same order as adatom attachment barriers need to be used. In this scenario, the only property of the system that makes tetramer attachment a predominant effect in graphene growth is the larger diffusion of tetramers as compared to adatoms. In the next section we are going to confirm that under these circumstances tetramer attachment is still the dominant effect.

\section{Error analysis: island growth velocity}

In this section we present an error analysis on the KMC model described in this thesis. Statistical errors due to the stochastic nature of the method are already discussed in the context of the standard model above, and that discussion is not repeated here. Instead, we concentrate in the evaluation of the model used and its computational implementation. We look at the velocity of island growth and find that the KMC model is consistent with the $m$-cluster model proposed experimentally for the growth of epitaxial graphene. %Second, we compare the KMC results with a RE analysis and find qualitative agreement between the two.

%\subsection{Island growth velocity}

%In this subsection we present the analysis of the island growth velocity for the proposed KMC model. 

In Ref.\cite{ref:loginova2008} the island growth velocity is found to be a nonlinear function of the adatom concentration, ultimately leading to the proposal of a growth model for graphene with a 5-cluster intermediate species. As discussed in Section \ref{subsec:exp}, postulating that the energy barrier for adatom attachment to islands is larger than the energy barriers for clusters of size $m$ to form and later to join islands, it can be shown that the island growth velocity obeys

\begin{equation}
v = B\left[\left(\frac{n}{n_{eq}}\right)^m-1\right] \label{eq:powerlaw}
\end{equation}
where $n$ is the adatom concentration, $n_{eq}$ is the adatom concentration at equilibrium and $B$ is a temperature dependent constant. Experimentally, it is found that $m \simeq$ 5, and the intermediate species are clusters formed by 5 carbon atoms. In our KMC model, where tetramers play the role of the intermediate species, we expect to have a power law with exponent $m \simeq$ 4 in Eq.(\ref{eq:powerlaw}). This result will provide a consistency check for the model used in this work.

The island growth velocity $v$ is defined as \cite{ref:loginova2008}

\begin{equation}
v = \frac{1}{P}\frac{dA}{dt}
\end{equation}
for island perimeter $P$ and area $A$. For simplicity, we consider circular islands with $P = 2\pi R$ and $A = \pi R^2$ for radius $R$. The circular island approximation is expected to be accurate at high temperatures where we have seen that the system tends to large circular islands trying to minimise the edge free energy. With a total area $A = N$ for $N$ carbon atoms in the island and an island radius $R = \sqrt{N/\pi}$, then 

\begin{equation}
P = 2\sqrt{\pi N} \hspace{1cm} \mbox{and} \hspace{1cm} A = N.
\end{equation}
The estimate for the island growth velocity $v(t)$ at time $t$ in the circular island approximation is then

\begin{equation}
v(t) \simeq \frac{1}{P}\frac{\Delta A}{\Delta t} \simeq \frac{1}{2\sqrt{\pi N(t)}}\frac{N(t +\Delta t)-N(t)}{\Delta t}.
\end{equation}

\begin{figure}[h!]
\centering
\subfloat[$E_q = 0.10$ eV]{\includegraphics[width=0.45\textwidth]{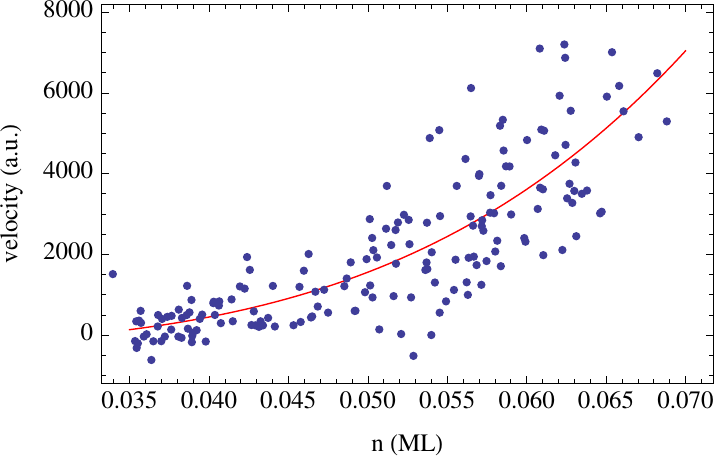}}
\hspace{0.07cm}
\subfloat[$E_q = 0.30$ eV]{\includegraphics[width=0.45\textwidth]{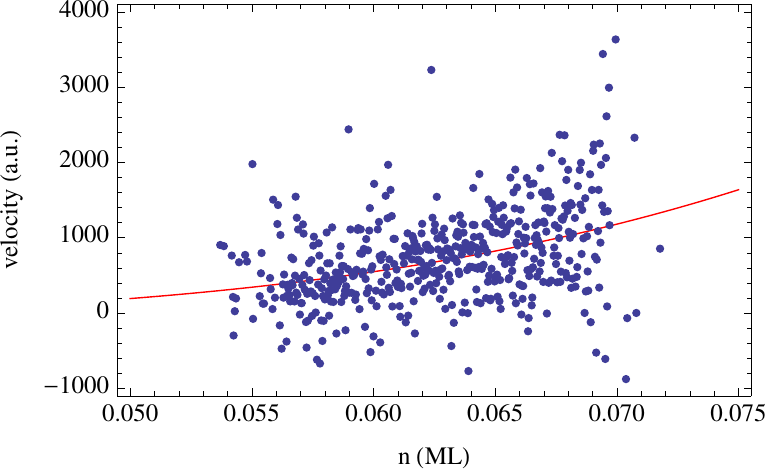}}\\
\caption{\label{fig:velocity}Island growth velocity as a function of adatom coverage at $T =$ 700K for tetramer attachment barriers (a) $E_q = 0.10$ eV and (b) $E_q = 0.30$ eV. A fit to Eq.(\ref{eq:powerlaw}) is also shown.}
\end{figure}

In Figure \ref{fig:velocity} we plot the $v$ as a function of $n$ for tetramer attachment barriers $E_q = 0.10, 0.30$ eV. The other parameters are those in Table \ref{tab:gparameters} and $E_k = 1.40$ eV, $E_p = 0.30$ eV. The data corresponds to $T = 700$ K, the large temperature limit discussed above, and it is then found that $n_{eq} = 0.032$ ML for $E_q = 0.10$ eV and $n_{eq} = 0.039$ ML for $E_q = 0.30$ eV. 

\begin{table}[h!]
\centering
\begin{tabular}{lll}
\hline
 & $E_q = 0.10$ eV & $E_q = 0.30$ eV\\ 
\hline
$B$ & $299.6 \pm 88.3$ & $94.9 \pm 43.7$ \\
$m$ & $4.09 \pm 0.43$ & $4.44 \pm 0.80$ \\
\hline
\end{tabular}
\caption{Least-squares fit values of the parameters $B$ and $m$ in Eq.(\ref{eq:powerlaw}) for tetramer attachment barriers $E_q = 0.10$ eV and $E_q = 0.30$ eV. \label{tab:fitparameters}}
\end{table}

The fit has been done using the Mathematica package with the model given by Eq.(\ref{eq:powerlaw}) and leaving as free parameters to fit $B$ and $m$. The least-squares fit parameters are shown in Table \ref{tab:fitparameters} where the quoted errors are the standard error. We note that the fit parameter $m$ is highly dependent on the value of $n_{eq}$ which needs to be determined with high accuracy. 

In both cases the fit value for $m$ indicates that the implementation of the model is consistent with the $m$-cluster model proposed by Loginova and co-workers, in our case with $m\simeq4$ because we used tetramers as the intermediate species. In the model described by Eq.(\ref{eq:powerlaw}), the attachment to graphene is via $m$-clusters. In the present study, the growth is via both tetramers and adatoms so we expect some discrepancies with Eq.(\ref{eq:powerlaw}). For $E_q = 0.10$ eV, $m$-cluster attachment is more predominant over adatoms than for the case with $E_q = 0.30$ eV, and indeed we observe a better fit. However, the discussion above in Subsection \ref{subsec:Neq} about $N_{eq}$ suggests that the appropriate value for tetramer attachment barrier is $E_q = 0.30$ eV. Even though the fit is not as accurate as for $E = 0.10$ eV, the velocity analysis here suggests that $E_q = 0.30$ eV is still consistent with Eq.(\ref{eq:powerlaw}). This indicates that the larger tetramer diffusivity still makes this species the predominant one in graphene growth.

Computationally, island sizes at time intervals of $\Delta t = 2.5$ ms have been used to obtain the results shown in Fig. \ref{fig:velocity}. Large islands can break-up or coalesce during growth, and then $v$ is not evaluated as the values do not correspond to growth by attachment of adatoms or tetramers only. Also, a deposition flux $F = 10$ ML/s has been used to explore the larger values of the density $n$.

Different approaches to check the validity of the model and its implementation can be considered outside of the results obtained within the model. In Appendix \ref{app:re} we present a RE comparison with the present KMC model that further validates it.

\chapter{Conclusions}\label{ch:conclusions}

\section{Summary}

In this thesis we have introduced a KMC model to study the physical processes governing the kinetics of epitaxial graphene growth on metal substrates. All the experimental observations have been explained within the model, and the relevant kinetic paths determining them have been identified.

The behaviour of adatom density at the onset of nucleation has been explained by the inclusion of both a large critical island size and a tetramer break-up rate. These conclusions are in agreement with previous RE studies.

The adatom density at equilibrium has been found to be highly dependent on the adatom attachment barrier to islands. With no such barrier, the adatom density at equilibrium is vanishingly small, but tuning the energetics of such a process can lead to the desired concentration of adatoms at equilibrium.

The island density at equilibrium has been found to be dependent on the attachment barrier of tetramers to islands. Without such barrier, few nucleations occur because tetramers are more likely to attach to an existing island than to nucleate new islands. Only with the inclusion of a large barrier the desired behaviour in the island density at equilibrium is obtained. This means that the dominance of tetramers in graphene growth is via their larger diffusivity rather than due to larger attachment barriers for adatoms than for tetramers.

%We also note that the KMC model presented here to study epitaxial growth is the first to incorporate an itermediate species in the system. The algorithm is not drastically changed, but the definition of the behaviour of the new species together with the rates associated with the new processes need to be added. 

\section{Further work}

The present work has allowed the determination of the relevant physical processes governing the particular behaviour of the kinetics of epitaxial graphene growth. The next natural step is to use this knowledge to construct an accurate KMC model to be compared directly with experiment. First, the various energy barriers in the problem need to be determined. A standard way of accomplishing this is by electronic structure calculations using well-established first-principles methods such as density functional theory. Also, both the metal substrate (lattice) and the graphene structure need to be hexagonal. Finally, an exact intermediate 5-cluster species instead of the more convenient tetramers needs to be included. Difficulties in this model might arise when comparing with experiment because high temperatures and large time intervals need to be studied, and this might be computationally too demanding. The code can be further optimised, but some approximations might still be necessary to reach the scales of interest.

The detemination of the relevant energy barriers for epitaxial graphene growth by means of \textit{ab initio} techniques can also be used in similar studies such as lattice-free KMC.

%Improve KMC algorithm in various directions to try to reach experimentally relevant time scales.

%Maybe talk about Vvedensky's proposed PhD project using lattice free KMC.

\vspace{0.7cm}

\noindent \textbf{Words in text: 9986.}

\appendix
\chapter{Island growth model} \label{app:growth}

In this appendix we consider the role of islands in the KMC model presented in Chapter \ref{ch:tetramer}.

The model presented in this thesis falls within the generic model class for the treatment of islands, but the rules on the existence of tetramers and nucleation of islands make it different from a pure bond-counting model. Generic models \cite{ref:evans2006} are those with local rules such as bond counting according to which the system evolves, and are different from the so-called tailored models. In the latter, some specifications on the system evolution are made based on the knowledge already existing about the system. For instance, a critical island size can be defined above which attachment of adatoms to islands is irreversible. Another possibility is the \textit{prescribed island growth sequence} model \cite{ref:evans2006} that we will discuss here in the context of graphene.

In a prescribed island growth sequence model, adatoms that attach to an island are instantenously moved to a prescribed position on the island independently of the attachment position. This is equivalent to having an infinite edge diffusitivity to the most stable sites. It is used in systems known to grow in certain shapes, for instance in growth on metal (100) where the islands are known to be nearly squares and each new attached adatom is immediately incorporated in the sequence to give the desired square shape.

\begin{figure}[h!]
\centering
\includegraphics[scale=0.3]{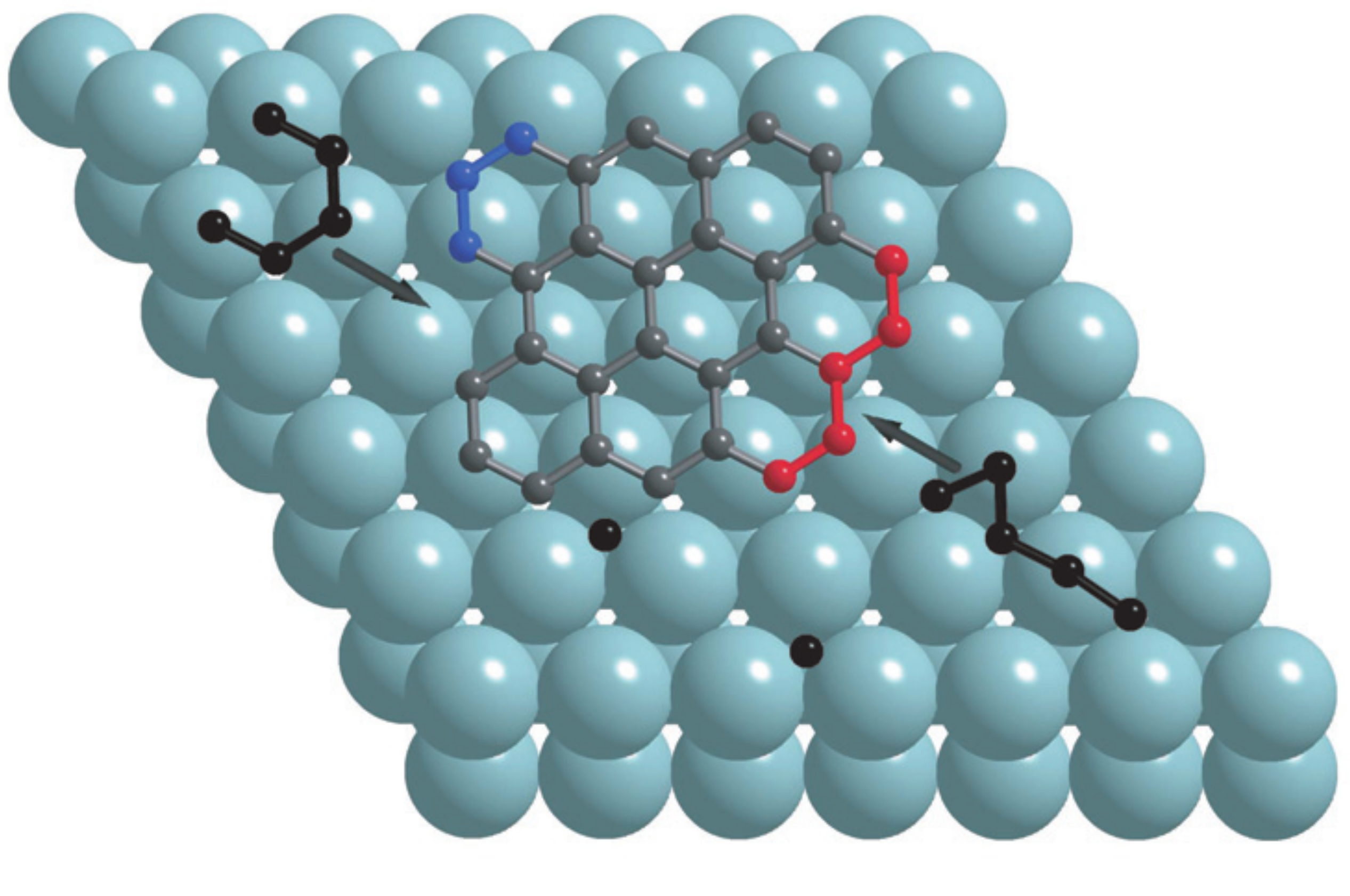}
\caption{\label{fig:attachment}Schematic of 5-cluster attachment to islands. Taken from \cite{ref:loginova2008}}
\end{figure}

For graphene growth via 5-clusters, the structure of the hexagonal lattice makes the 5-clusters have the appropriate size to represent an attachment in which growth only occurs via the formation of new complete hexagons. In Fig. \ref{fig:attachment} we can see such configurations schematically, where we can note that 3-clusters would also lead to such hexagon by hexagon growth and 5-clusters actually lead to double hexagon growth. This indicates that graphene grows in a very prescribed manner, and the tetramer attachment and bond counting scheme used in this thesis might not be entirely accurate in describing it. A prescribed island growth sequence, by which tetramer attachment leads to an immediate relaxation into a more stable configuration might be more appropriate. 

\begin{figure}[h!]
\centering
\includegraphics[scale=0.3]{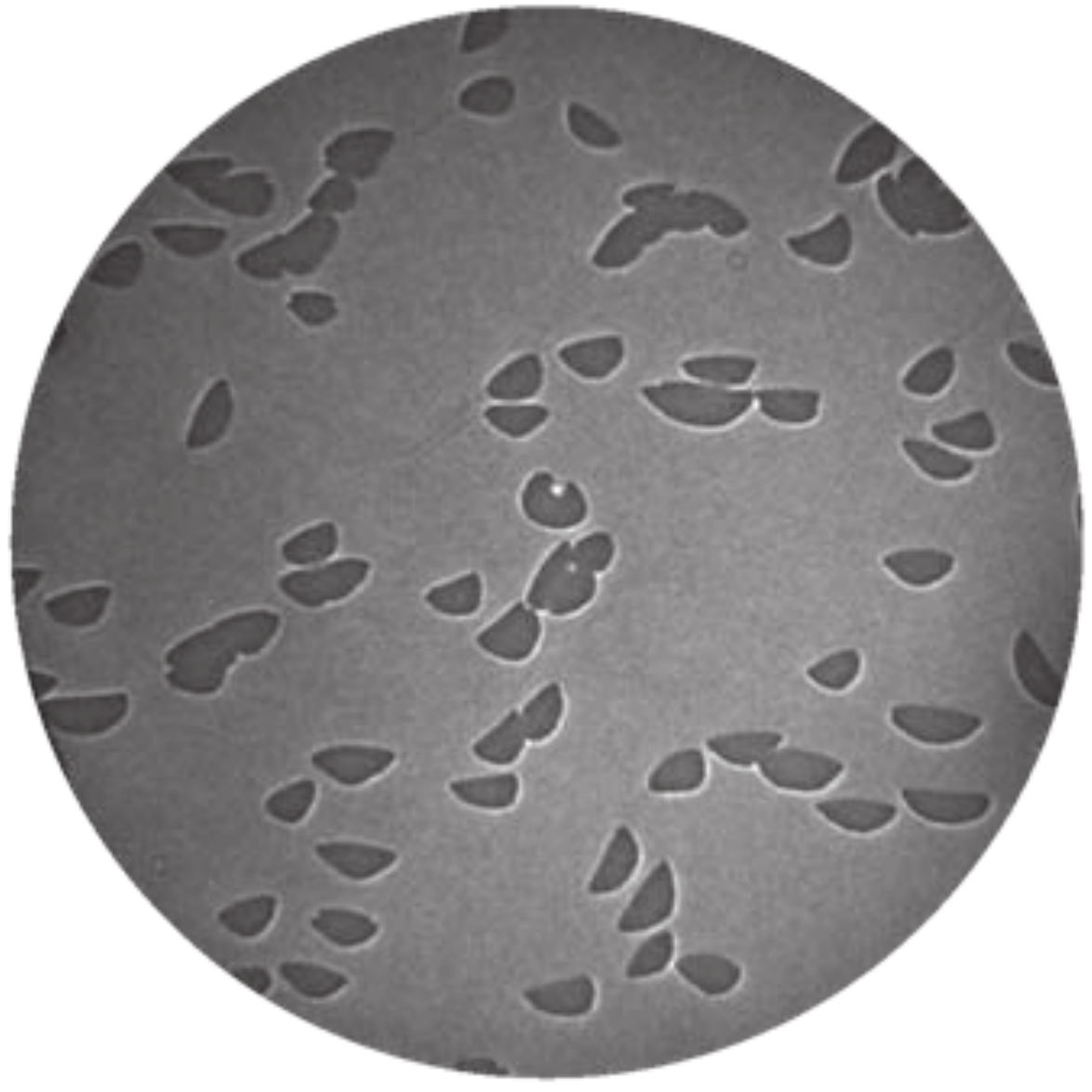}
\caption{\label{fig:graphene}LEEM image of a Ru(0001) substrate with grown graphene islands on it. Taken from \cite{ref:loginova2008}}
\end{figure}

The island shape observed in graphene and shown in Fig. \ref{fig:graphene} also suggests a very specific growth sequence, possibly the result of this attachment mechanism.

\chapter{Island size distribution} \label{app:isd}

\begin{figure}[h!]
\centering
\includegraphics[scale=1.3]{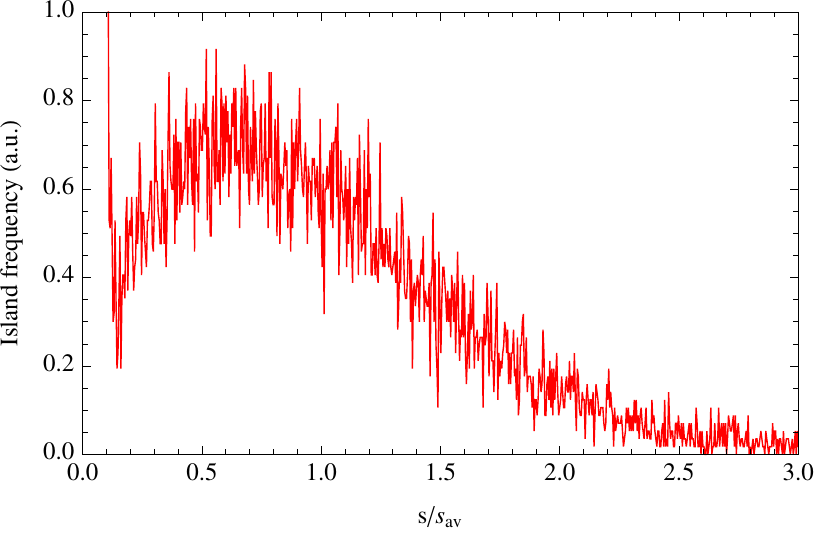}
\caption{\label{fig:appendixdistribution}Island size distribution for the graphene KMC model at $T = 500$ K.}
\end{figure}

In this appendix we present the preliminary data for the island size distribution of the KMC model for the description of epitaxial growth. For the reasons discussed in Chapter \ref{ch:standard}, data for island size distribution is very noisy, and to get significant results many instances of the system need to be used. Time constraints have limited the results to the case of $T = 500$ K for the graphene model so far. These are shown in Fig. \ref{fig:appendixdistribution} for the parameters in Table \ref{tab:gparameters} and $E_k = 1.40$ eV, $E_p = 0.30$ eV and $E_q = 0.30$ eV as usual. The shape observed is to be expected for the island size distribution. However, any detailed analysis will require the behaviour of island size distribution with temperature.

\chapter{Rate equations and kinetic Monte Carlo comparison} \label{app:re}

In this appendix we compare the KMC model presented in this thesis with the RE approach by Zangwill and Vvedensky \cite{ref:zangwill}. The objective of the comparison is to validate the model used in this work.

\begin{table}[h!]
\centering
\begin{tabular}{lll}
\hline
 & RE & KMC\\ 
\hline
$E_a$ & 1.00 eV & 1.00 eV \\
$E_t$ & 0.80 eV & 0.80 eV \\
$E_k$ & 1.40 eV & 1.40 eV \\
$E_b$ & -- & 0.20 eV \\
$E_i$ & 1.60 eV & --  \\
\hline
\end{tabular}
\caption{Least-squares fit values of the parameters $B$ and $m$ in Eq.(\ref{eq:powerlaw}) for tetramer attachment barriers $E_q = 0.10$ eV and $E_q = 0.30$ eV. \label{tab:rekmc}}
\end{table}

The rate equations model used is the one presented by Zangwill and Vvedensky \cite{ref:zangwill} as described in Sectin \ref{subsec:theo} above. This RE model includes parameters to describe adatom and tetramer diffusion, tetramer break-up and adatom detachment $E_i$ from islands. The KMC model has corresponding parameters for the first three quantities, but for the latter it is the adatom bond energy between adatoms in an island that maps the effect. In order to carry out the comparison, RE have been used with $i = 4$ and $j = 6$ and KMC has been used with $E_p = E_q = 0$. The rest of parameters are shown in Table \ref{tab:rekmc}. A temperature of $T = 650$ K is used with a flux $F = 1$ ML/s. The total coverage reached is $\theta = 0.2$ ML corresponding to 0.2 seconds, and the system is allowed to relax for a further 0.3 seconds.

\begin{figure}[h!]
\centering
\subfloat[Adatom density]{\includegraphics[width=0.47\textwidth]{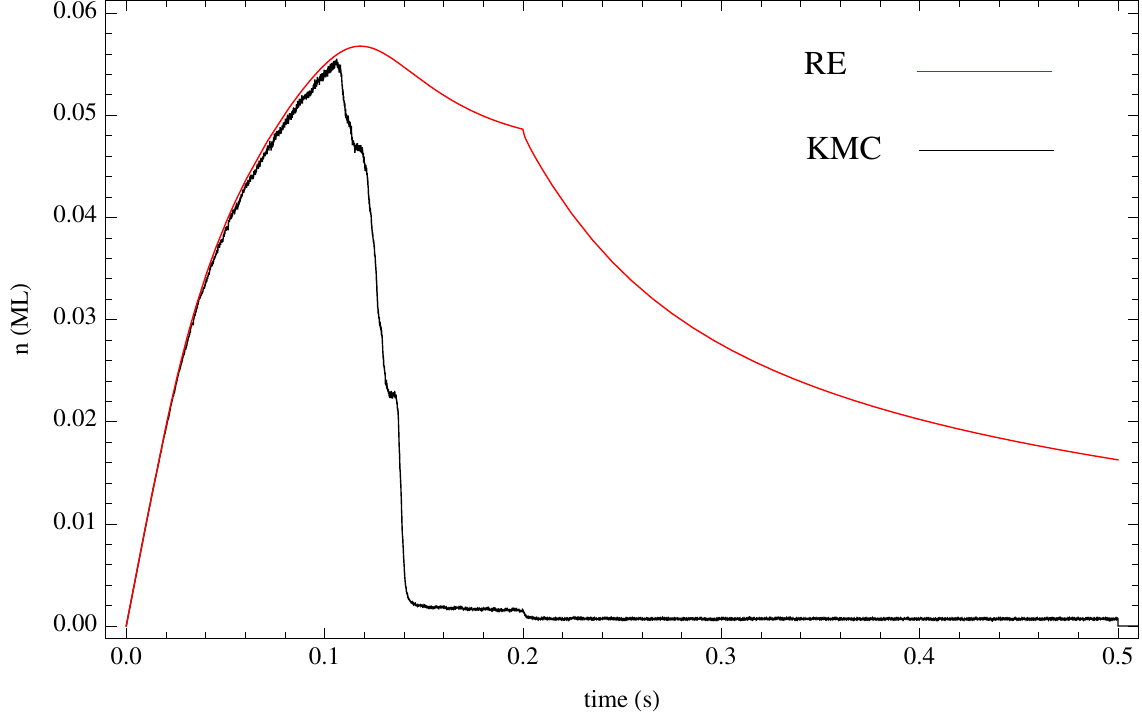}}
\hspace{0.07cm}
\subfloat[Tetramer density]{\includegraphics[width=0.47\textwidth]{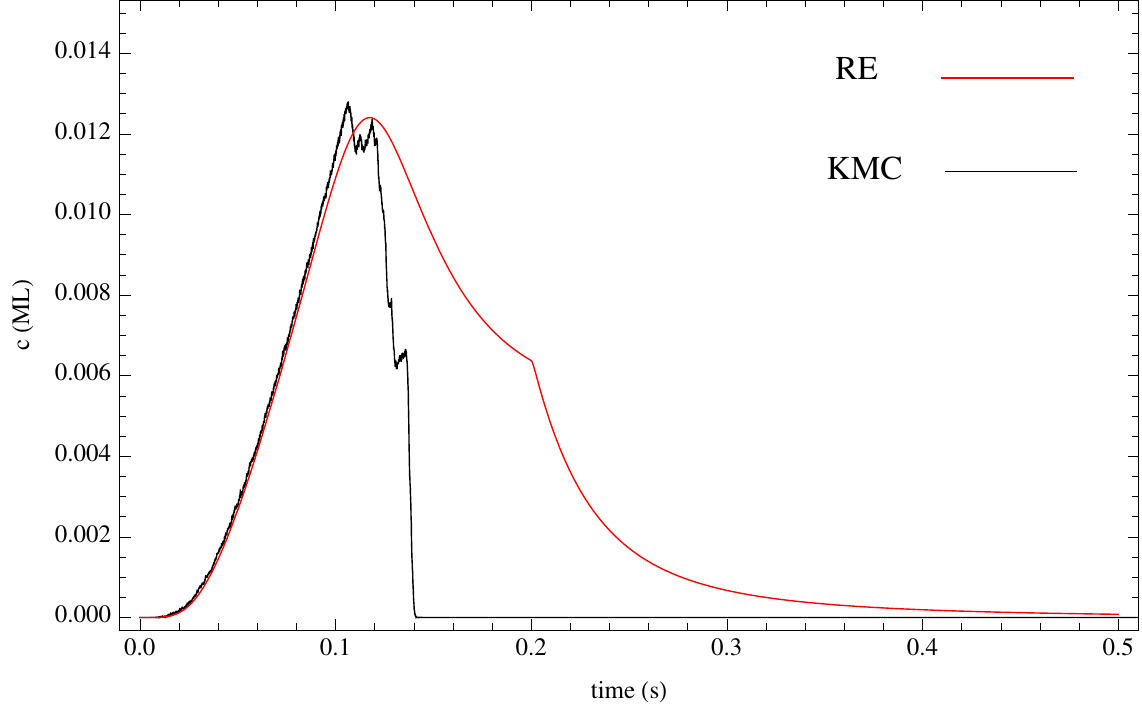}}\\
\caption{\label{fig:recomparison}Comparison between RE and KMC model for (a) adatom density and (b) tetramer density.}
\end{figure}

Figure \ref{fig:recomparison} shows the adatom concentration $n$ and the tetramer concentration $c$ as a function of temperature. In both cases, KMC and RE coincide up to the nucleation point at about $t\sim0.1$~s, but a clear disagreement appears after this. The RE overestimate the concentration of both adatoms and tetramers. In the KMC approach, islands have spatial extent, so their adatom and tetramer capture zones are large. However, in the RE the islands are point-like, and their capture zones are reduced, thus leading to a smaller decrease in the concentration of adatoms and tetramers.

\begin{figure}[h!]
\centering
\includegraphics[scale=1.0]{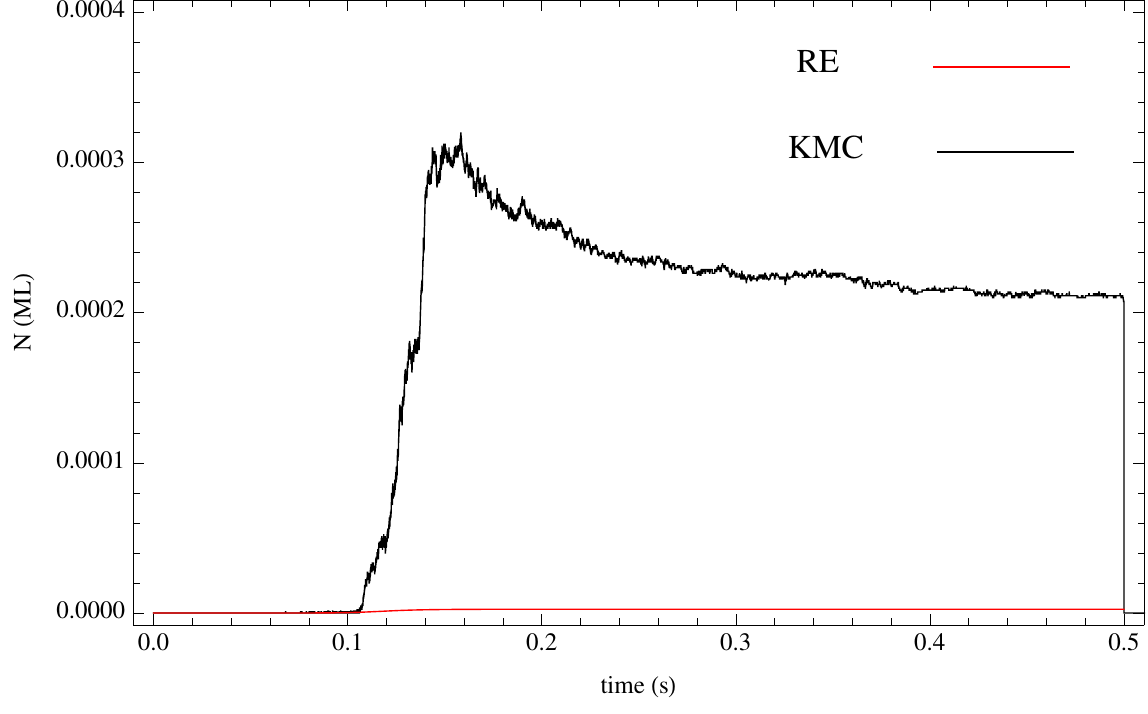}
\caption{\label{fig:reislandcomparison}Comparison between RE and KMC model for the island density.}
\end{figure}

In Fig. \ref{fig:reislandcomparison} we show the comparison between RE and KMC for the island density. Unlike the above, this quantity is badly reproduced by the rate equations as already discussed in the work by Zangwill and Vvedensky. The island concetration is underestimated by the rate equations, and again this is attributed due to the lack of spatial information \cite{ref:zangwill}.

\bibliographystyle{h-physrev}%to order them with order of appearance
\bibliography{graphene.bib}

\end{document}